\documentclass[%
 reprint,
superscriptaddress,
 amsmath,amssymb,
 aps,
]{revtex4-1}

\usepackage{graphicx}
\usepackage{dcolumn}
\usepackage{bm}
\usepackage{tabularx}
\usepackage[T1]{fontenc}
\usepackage[utf8]{inputenc}
\usepackage{xcolor}
\usepackage{amsmath}

\begin{document}

\title{Numerical Solver for the out-of-equilibrium time dependent Boltzmann Collision operator: Application to 2D materials}

\author{I.~Wadgaonkar}
\affiliation {Nanyang Technological University, 21 Nanyang Link, Singapore, Singapore}
\author{M.~Wais}
\affiliation {Nanyang Technological University, 21 Nanyang Link, Singapore, Singapore}
\affiliation {Technical University Wein, Vienna, Austria}
\author{M.~Battiato}
\email{marco.battiato@ntu.edu.sg}
\affiliation {Nanyang Technological University, 21 Nanyang Link, Singapore, Singapore}
 \date{\today}

\begin{abstract}
The Time Dependent Boltzmann equation (TDBE) is a viable option to study strongly out-of-equilibrium thermalization dynamics which are becoming increasingly critical for many novel physical applications like Ultrafast thermalization, Terahertz radiation etc. However its applicability is greatly limited by the impractical scaling of the solution to its scattering integral term. In our previous work\cite{Michael} we had proposed a numerical solver to calculate the scattering integral term in the TDBE and then improved on it\cite{1DPaper} to include second degree momentum discretisation and adaptive time stepping.  Our solver requires no close-to-equilibrium assumptions and can work with realistic band structures and scattering amplitudes. Moreover, it is numerically efficient and extremely robust against inherent numerical instabilities. While in our previous work \cite{1DPaper} we showcased the application of our solver to 1D materials, here we showcase its applications to a simple 2D system and analyse thermalisations of the introduced out-of-equilibrium excitations. The excitations added at higher energies were found to thermalise faster than those introduced at relatively lower energies. Also, we conclude that the thermalisation of the out-of-equilibrium population to equilibrium values is not a simple exponential decay but rather a non-trivial function of time. Nonetheless, by fitting a double exponential function to the decay of the out-of-equilibrium population with time we were able to generate quantitative insights into the time scales involved in the thermalisations.
\end{abstract}

\pacs{Valid PACS appear here}
\maketitle


\section{\label{sec:level1}Introduction}

The discovery of a number of novel effects in low dimensional materials \cite{Beaurepaire1996,Terada2008,Crepaldi2012,Garcia2011,Cacho2015,Battiato2018PRLHalfMetals,Cheng2019,Wang2018THzTopolInsul,Lindemann2019UltrafastSpinLasers} has led to an increasing interest in femtosecond dynamics in recent times. Although the experimental investigations continue to probe deeper into the underlying physics \cite{Fujimoto1984,Elsayed1987,Schoenlein1987,Brorson1990,Fann1992,Hertel1996,Fatti2000,Stamm2007}, the theoretical and numerical efforts in this direction have been facing significant challenges. The foremost challenge is to effectively describe the intermingling between laser interactions and scatterings and transport of various quasi particles \cite{malinowski2008control, battiato2010superdiffusive,rudolf2012ultrafast,kampfrath2013terahertz, eschenlohr2013ultrafast, battiato2016ultrafast,Freyse2018}. Moreover, since most of the interesting effects occur in the far-out-of-equilibrium regime \cite{Bagsican2020THzCNT}, the full description of the time varying out-of-equilibrium population becomes unavoidable adding to the complexity.

The Time Dependent Boltzmann equation (TDBE)  \cite{Boltzmann1872} has proven to be a viable alternative in handling this complexity and has already been successfully employed to describe gases, plasma and semiconductor's physics \cite{Snoke2011,colonna2016plasma,villani2002review,kremer2010introduction,saint2009hydrodynamic,colussi2015undamped,shomali2017monte,xu2017lattice}. Although the TDBE  discards the phase information and only accounts for the average occupation number of the quantum states, it has, nevertheless, been proved to be successful in preserving the leading order dynamic properties of interest. Moreover, in addition to being significantly cheaper than the fully quantum mechanical methods, the TDBE also provides a seamless integration between transport and scattering between different types of quasi particles \cite{Semiconductor1969,cercignani1988boltzmann,Abdallah1996,Mahan2000,Choquet2000,Rethfield2002,Majorana2004,Caceres2006,Tani2012,Vatsal2017,wais2018quantum}. 

Yet, however, there is a final hurdle in extending the application of the TDBE to the more interesting cases. While the transport part of the TDBE has been handled well \cite{morgan1990elendif,nabovati2011lattice,sellan2010cross,hamian2015finite,romano2015dsmc,Majorana2019BCTGraph,heath2012discontinuous,cockburn2001runge,li1998analytic,choquet2003energy,majorana2004charge,Singh2020Boltztransp,Singh2020Boltztransp2}, the scattering part, which becomes more significant in the out-of-equilibrium regime, has always been more challenging \cite{vangessel2018review,chernatynskiy2010evaluation,broido2007intrinsic,wu2013deterministic,bird1994molecular,homolle2007low,tcheremissine2006solution,ibragimov2002numerical,gamba2009spectral,pareschi2000numerical,mouhot2006fast,tani2012ultrafast,maldonado2017theory,ziman2001electrons,fischetti1988monte}. For classical particles, like gases and plasma, or in the close-to-equilibrium regime the scattering term can be simplified to a linear or at most quadratic operator of the particle's populations, which can be handled with today's computational resources. However, in the far-out-of-equilibrium regime the quantum statistics of the particle cannot be neglected and so the scattering term becomes, for eg. in the case of electron-electron collisions, a quartic operator owing to the presence of the Pauli factors. This indispensable complexity  of the scattering term is difficult to address using straightforward numerical methods and has been historically dealt with by introducing close-to-equilibrium assumptions which trade off the complexity of handling the full scattering term for limitations on the applicable range of the TDBE. Further, a reliable time propagation of the population requires that the calculation of the multidimensional integral of the extremely discontinuous integrands in the scattering term, be able to exactly conserve critical quantities like particle number, momentum and energy. Unfortunately, the standard numerical algorithms to calculate these multidimensional integrals fail in the conservation of the critical quantities and hence cannot be employed reliably. These inherent complexities of the TDBE must be resolved before it could find any real use in contributing to the numerical and theoretical efforts in femtosecond dynamics.

We have proposed a numerical algorithm\cite{Michael,Bagsican2020THzCNT} to calculate the scattering integral in the TDBE which requires no linearisation of the scattering operator, no close-to-equilibrium approximations, preserves particle,momentum and energy and can work with realistic dispersions and scattering amplitudes. We further extended it to include second order momentum discretisation and adaptive time stepping \cite{1DPaper}, which addressed the problem of numerical errors leading to locally unphysical values of the population. While in our previous work we applied the improved numerical solver to 1D materials, in this work we showcase the applications of our solver to 2D materials. In particular, we analyse the thermalization of electronic excitations in doped graphene beyond any close-to-equilibrium approximation.

\section{\label{sec:numericalmethod} The Boltzmann Equation: Collision Operator}

The collision operator in the TDBE \cite{Snoke2011,snoke2020solid} provides the time, $t$, evolution due to scattering events of the momentum resolved population $f_n(t,\vec{\bold{k}})$ of any quasiparticle in crystals, where the index $n$ includes both the quasiparticle type and the band label, while $\vec{\bold{k}}$ stands for the crystal momentum. The collision operator is the sum of collision integrals due to the possible scattering channels. Each combination of quasiparticles and bands will contribute one scattering integral. 

The form of the scattering integral depends on the number and the statistics of the involved states. For instance, a typical fermion-fermion four leg (i.e.~involving four states) scattering integral is written as 
\begin{widetext}
\begin{equation}\label{ScatteringIntegral}
\begin{split}
\Bigg(\frac{\partial f_0}{\partial t}\Bigg)_{\substack{n_0+ n_1\\ \Longleftrightarrow \\ n_2 + n_3}}= \sum_{\vec{\bold{G}}}  \iiint_{V^3_{BZ}} & d \vec{\bold{k}}_1\; d\vec{\bold{k}}_2\; d\vec{\bold{k}}_3 \; \;w^{e-e}_{0123}\left(\vec{\bold{k}}_0,\vec{\bold{k}}_1,\vec{\bold{k}}_2,\vec{\bold{k}}_3 \right)  \; \delta(\epsilon_{n_0}(\vec{\bold{k}}_0)+\epsilon_{n_1}(\vec{\bold{k}}_1)-\epsilon_{n_2}(\vec{\bold{k}}_2)-\epsilon_{n_3}(\vec{\bold{k}}_3)) \\ & \delta(\vec{\bold{k}}_0+\vec{\bold{k}}_1-\vec{\bold{k}}_2-\vec{\bold{k}}_3+\vec{\bold{G}}) \underbrace{[(1-f_0)(1-f_1)f_2 f_3-f_0f_1(1-f_2)(1-f_3)]}_{P_{0123}}
\end{split}
\end{equation}
\end{widetext}
Here, $f_0$ is short hand for $f_{n_0} (t,\vec{\bold{k}}_0)$  while $\epsilon_n(\vec{\bold{k}})$  is the quasiparticle dispersion which is assumed to be known. $w^{e-e}_{0123}$, which we call scattering amplitude, is, in principle, a function of the momenta $\vec{\bold{k}}_n$ of all the involved states in the scattering process and again assumed to be known by other means (or appropriately approximated). $\vec{\bold{G}}$ is the reciprocal lattice vector and the summation over $\vec{\bold{G}}$ in Eq.~\eqref{ScatteringIntegral} is required to account for umklapp scatterings. The triple integral over the volume of Brillouin zone, $V_{BZ}$, in conjunction with the two Dirac deltas ensures that all combinations of momenta, which satisfy energy and momentum conservation (up to a reciprocal lattice vector $\vec{\bold{G}}$), are accounted for. The phase space factor, denoted as $P_{0123}$, accounts for fermionic Pauli factors (yet it is easily generalisable to bosons) and includes the time reversed process.

Eq.~\eqref{ScatteringIntegral} can be easily generalised to quasiparticles with different statistics, scatterings involving different quasiparticles, as well as different number of legs (i.e. involved states), yet maintaining the same mathematical structure. The great advantage is that a numerical method developed for Eq.~\eqref{ScatteringIntegral} can be extended to its generalisation. These scattering integrals of the type in Eq.~\eqref{ScatteringIntegral} are typically approximated when computing time propagations \cite{battiato2010superdiffusive,battiato2016ultrafast,Snoke2011,Rethfield2002,jhalani2017ultrafast,wais2018quantum,Mahan2000,Snoke2007,maldonado2017theory, Sadasivam2017,Sanchez-Barriga2017,Battiato2018PRLHalfMetals,Kurosawa1971,Jacoboni1983,fischetti1988monte,Medvedev2011,Mattei2017,Nenno2018}. 

 The most common approximation is the close-to-equilibrium approximation \cite{maldonado2017theory,Rethfield2002, wais2018quantum, Sadasivam2017, Sanchez-Barriga2017, Battiato2018PRLHalfMetals, Kurosawa1971, Medvedev2011, Mattei2017}.For systems in contact with a thermal bath this condition is automatically satisfied \cite{Jacoboni1983}. Notice that in the close-to-equlibrium regime one or more populations from the phase factor, $P_{0123}$, in eq.\ref{ScatteringIntegral} is/are no longer a function of time but rather an equilibrium distribution. This can greatly simplify the scattering integral term. For eg. in a four leg electron-electron scattering process when the population in three legs can be assumed to be independent of time, they can be combined in a single parameter called 'relaxation time' which simplifies Eq.\ref{ScatteringIntegral} to a linear operator instead of a quartic one \cite{battiato2010superdiffusive,battiato2016ultrafast, jhalani2017ultrafast, fischetti1988monte, Nenno2018}. This approximation is commonly called the 'Relaxation time Approximation (RTA)'.We highlight that these approximations severely limit the applicability of the Boltzmann equation.

The solution of the full Eq.~\eqref{ScatteringIntegral} without approximations presents serious computational challenges. i) As all the populations appearing inside the high dimensional integral in Eq.~\ref{ScatteringIntegral} are time dependent and known only at run time, making it unavoidable to treat the integral as a quartic operator (note that assuming some of the involved populations as known and/or at equilibrium, immediately dramatically lowers the numerical complexity). This results in an impractical scaling of the computational and the storage cost of the discretised form of the scattering integral. ii) The presence of multiple Dirac deltas (depending on the dimensionality of the system considered), one of which has a highly non-trivial form, make the integration domain highly discontinuous. As such, the use of integration methods like Monte Carlo Integration is unreliable (one approach used in the community is to broaden the Dirac deltas, but that leads to violations of the conservation laws; such violation can be acceptable is close to equilibrium calculations, but unacceptable in the description of thermalisation, see more below). iii) A discretisation method lacking exact conservation of extensive quantities like particle number, momentum and energy prevents the computation of the full thermalisation dynamics.

 \section{Discretization}

In this work we aim at extending the second order approach developed in Ref.~\cite{1DPaper} to the 2D case. We will show that while the discretisation itself (basis function, projection, and structure of the scattering tensor) can be generalised fairly straightforwardly, the integration and the inversion of the Dirac deltas requires importantly more effort. 

First, for the sake of completeness and for the reader's convenience, we briefly summarise the structure of the method which is independent of the basis functions. We then describe the required modifications in the calculation of the scattering tensor elements owing to the second degree polynomial basis functions. For the sake of shortness, we focus on a single 2D band. However all the details are individually applicable to any number of bands and a generic type of scattering channel.  

The band and momentum resolved population distribution, $f_n(t,\vec{\bold{k}})$, and dispersion relation, $\epsilon_n(\vec{\bold{k}})$ (which is assumed to be known) are assumed to be defined over a domain which can be a compact subset of the Brillouin zone. Notice that each band can have its own domain (this allows for the exclusion from the calculations of areas of the Brillouin zone not involved in the dynamics).

\subsection{Basis functions}

First, we split the domain of the band into non-overlapping elements thereby forming a mesh. For 2D system we use a mesh consisting of equal rectangular elements (but the approach can be equally applied to triangular elements).  We project the solution and the dispersion on a set of band $n$ specific momentum basis functions \(\Psi_{\substack {Aa \\n}}(\vec{\bold{k}})\). The basis functions are zero everywhere in the domain except on the element identified by the index $A$ and they are continuous everywhere except on the edges of the element $A$. Each element in the mesh can have several basis functions which are labelled by sub index $a$, such that the basis functions are linearly independent. In this study we assume that the basis functions are orthonormal:
\begin{equation}\label{Orthonorm}
    \int d^2\vec{\bold{k}} \;\; \Psi_{\substack {Bb \\n}}(\vec{\bold{k}}) \;\; \Psi_{\substack {Aa \\n}}(\vec{\bold{k}}) = \delta_{AB} \delta_{ab}
\end{equation}
where $\delta$ is the Kronecker delta. The method is equally valid for a non-orthonormal basis set with the RHS in Eq.~\ref{Orthonorm} replaced by a mass matrix. A few sample basis functions from the basis set chosen for this study are shown in fig.\ref{fig:Basisfnc}. The detailed derivation of the exact functional form of the chosen basis set is given in Appendix \ref{Appendix:BasisFunctions}.

\begin{figure}
\centering
\includegraphics[width=\columnwidth]{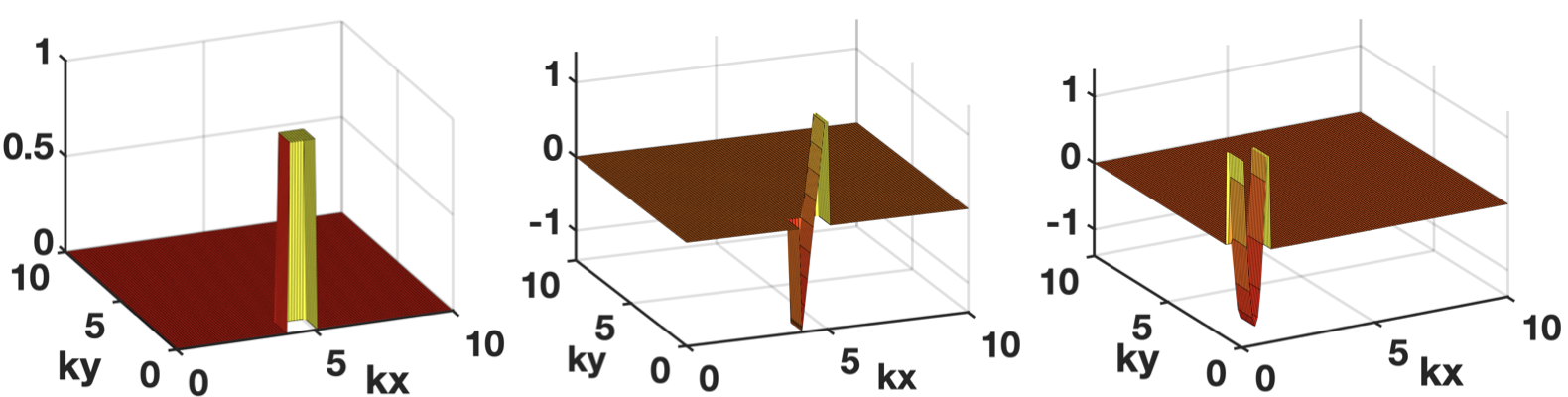}
\caption{ 
Sample chosen basis functions for a particular band n. From left: $\Psi_{\underset{n}{A0}} (\vec{\bold{k}})$ 
}
\label{fig:Basisfnc} 
\end{figure}

Once the basis functions are defined, we can express the solution, $f_n(t,\vec{\bold{k}})$ and the dispersion,$\epsilon_n(\vec{\bold{k}})$ as a linear combination of these basis functions:
\begin{equation}\label{fandEProjection}
\begin{split}
f_n(t,\vec{\bold{k}})=&\sum_{B b} f_{\substack{B b \\ n}}(t) \;\Psi_{\substack{B b \\ n}} (\vec{\bold{k}}) \\
\epsilon_n(\vec{\bold{k}})=&\sum_{B b} \epsilon_{\substack{B b \\ n}} \; \Psi_{\substack{B b \\ n}} (\vec{\bold{k}})\\
1=&\sum_{B b} 1_{\substack{B b \\ n}} \; \Psi_{\substack{B b \\ n}} (\vec{\bold{k}})
\end{split}
\end{equation}
where $f_{\substack{B b \\ n}}(t), \epsilon_{\substack{B b \\ n}}$ and $1_{\substack{B b \\ n}}$ are the coefficients of the discretised representations of the population distribution function, dispersion and the constant function respectively. Notice that the solution has been so far only partially discretised. Moreover we have, for later convenience, also written the discretized representation of a constant function which is equal to 1 on the whole k space. 

\subsection{Momentum discretisation}

Projecting Eq.~\eqref{ScatteringIntegral} on the chosen orthonormal basis functions, using Eq.~\ref{fandEProjection} and using the fact that the basis functions are non-zero only over a single element (see Ref.~\cite{Michael} for details) we get the final expression for the semi-discretised form of the time propagation as
\begin{widetext}
\begin{equation}\label{CollisionIntegral}
\begin{split}
\sum_{A a'} \frac{d f_{\substack{A a' \\n_0}}(t)}{d t}=\sum_{A a} \sum_{B b} \sum_{C c} \sum_{D d} S^{a' a b c d}_{\substack{A A B C D \\ n_0 n_1 n_2 n_3}} & \Bigg((1_{\substack{A a \\ n_0}}- f_{\substack{A a \\n_0}}(t)) (1_{\substack{B b \\ n_1}}- f_{\substack{B b \\n_1}}(t))  f_{\substack{C c \\n_2}}(t)f_{\substack{D d \\n_3}}(t) -\\
& f_{\substack{A a \\n_0}}(t) f_{\substack{B b \\n_1}}(t) (1_{\substack{C c \\ n_2}}-f_{\substack{C c \\n_2}}(t)) (1_{\substack{D d \\ n_3}}-f_{\substack{D d \\n_3}}(t)) \Bigg) \\
\end{split}
\end{equation}
with
\begin{equation}\label{ScatteringTensor}
\begin{split}
S^{a' a b c d}_{\substack{A A B C D \\ n_0 n_1 n_2 n_3}}=\sum_\bold{G} \int\displaylimits_{\substack{A \\ n_0}}\int\displaylimits_{\substack{B \\ n_1}}\int\displaylimits_{\substack{C \\ n_2}}\int\displaylimits_{\substack{D \\ n_3}} &  w^{e-e}_{0123} \;\; d^2 \vec{\bold{k}}_0 \;\; d^2 \vec{\bold{k}}_1 \; \;  d^2 \vec{\bold{k}}_2 \; \;  d^2 \vec{\bold{k}}_3 \; \; \Psi_{\substack{A a' \\ n_0}} (\vec{\bold{k}}_0) \; \Psi_{\substack{A a \\ n_0}} (\vec{\bold{k}}_0) \; \Psi_{\substack{B b \\ n_1}} (\vec{\bold{k}}_1) \; \Psi_{\substack{C c \\ n_2}} (\vec{\bold{k}}_2) \; \Psi_{\substack{D d \\ n_3}} (\vec{\bold{k}}_3)\\
& \delta(\vec{\bold{k}}_0+\vec{\bold{k}}_1-\vec{\bold{k}}_2-\vec{\bold{k}}_3+\bold{G}) \\ 
& \delta\Big(\sum_{\alpha} \epsilon_{\substack{A \alpha \\ n_0}} \; \Psi_{\substack{A \alpha \\ n_0}} (\vec{\bold{k}}_0)+\sum_{\beta} \epsilon_{\substack{B \beta \\ n_1}} \; \Psi_{\substack{B \beta \\ n_1}} (\vec{\bold{k}}_1)-\sum_{\gamma} \epsilon_{\substack{C \gamma \\ n_2}} \; \Psi_{\substack{C \gamma \\ n_2}} (\vec{\bold{k}}_2)-\sum_{\xi} \epsilon_{\substack{D \xi \\ n_3}} \; \Psi_{\substack{D \xi \\ n_3}} (\vec{\bold{k}}_3) \Big)
\end{split}
\end{equation}
\end{widetext}
Eq.\ref{CollisionIntegral} is still semi-discrete as the time variable has not been discretised yet. We refer to $S^{a' a b c d}_{\substack{A A B C D \\ n_0 n_1 n_2 n_3}}$ as the scattering tensor, and it contains all the information about the scattering. We highlight here that the integrals in Eq.~\ref{ScatteringTensor} are no longer over the whole Brillouin zone but rather only on single elements owing to our choice of piecewise continuous polynomials as basis functions. This has important consequences on the scaling of the overall computational cost. We study this aspect in further detail in sec.~\ref{scaling}.

\subsection{Time Propagation}

Once the scattering tensor is calculated, the population can be propagated in time, using Eq.~\ref{CollisionIntegral}, by contracting the scattering tensor with the instantaneous populations. The discretisation in time for Eq.~\ref{CollisionIntegral}, however, would require an intelligent choice of a suitable time propagation scheme given that multiple time scales are normally involved in the far-from-equilibrium thermalisation dynamics. Following the conclusions of our previous work, we choose to employ adaptive time stepping through DP853 numerical algorithm.\cite{1DPaper}

\section{Calculation of the scattering tensor elements}

In this section we detail the calculation of the scattering tensor elements using Eq.~\ref{ScatteringTensor} in the case of 2D materials. Some of the issues are similar to the 1D case, the most important of which is the presence of multiple Dirac deltas that makes the integration domain an extremely complex and discontinuous hyper-surface. However new challenges arise, as the number of Dirac deltas is increased. Moreover, while in the 1D case the Dirac deltas constrained completely the momenta of two legs, now one of the legs is only partially constrained.

Our choice of polynomial basis functions converts the expressions in the momentum Dirac deltas to polynomials of first degree, while in the expression of the energy in a second degree one. In spite of the increased dimensionality it is  possible to analytically invert all the Dirac deltas. 

We note that the argument of the momentum Dirac delta is not symmetric with respect to the legs. This is inconvenient since one would have to explicitly write analytic expressions for all the different choices of order of inversion. To avoid this issue, we first perform a mapping of variables as: $k_1 \rightarrow x_A ; k_2 \rightarrow x_B ; k_3 \rightarrow -x_C ; k_4 \rightarrow -x_D +G $. This mapping brings the momentum Dirac delta in a completely symmetric form as $\delta(x_A+x_B+x_C+x_D)$. Now we invert the energy Dirac delta with respect to the first two variables. Note that we can use the same integration routine for inverting the energy Dirac delta with respect to any couple of variables by simply altering the mapping. 

Let us stress that the mapping above, also affects the limits of integration, which depend on the elements involved. Appendix \ref{Appendix:Phi} details the inversion of Dirac deltas and the derivation of the final structure of the expression for Monte Carlo integration which ends up having the form as:

\begin{widetext}
\begin{equation}\label{MonteCarloIntegration}
\begin{split}
\Phi[&F[],l_{Ax},h_{Ax},l_{Ay},h_{Ay},l_{Bx},h_{Bx},l_{By},h_{By},l_{Cx},h_{Cx},l_{Cy},h_{Cy},l_{Dx},h_{Dx},l_{Dy},h_{Dy},\mu_0,\mu_{A1},\mu_{A2},\mu_{A3},\mu_{A4},\mu_{A5},\mu_{B1},\mu_{B2}\\
&,\mu_{B3},\mu_{B4},\mu_{B5},\mu_{C1},\mu_{C2},\mu_{C3},\mu_{C4},\mu_{C5},\mu_{D1},\mu_{D2},\mu_{D3},\mu_{D4},\mu_{D5}]\\
=&\int_{l_{By}}^{h_{By}}dy_B\int_{l_{Cx}}^{h_{Cx}}\int_{l_{Cy}}^{h_{Cy}}dx_Cdy_C\int_{l_{Dx}}^{h_{Dx}}\int_{l_{Dy}}^{h_{Dy}}dx_Ddy_D\\
&\Bigg(\frac{F[x_{A+}[y_B,x_C,y_C,x_D,y_D],y_A[y_B,y_C,y_D],x_{B+}[y_B,x_C,y_C,x_D,y_D],y_B,x_C,y_C,x_D,y_D]}{\sqrt{D[y_B,x_C,y_C,x_D,y_D]}}\\
&\Theta_{[l_{Ax},h_{Ax},l_{Ay},h_{Ay}]}[x_{A+}[y_B,x_C,y_C,x_D,y_D],y_A[y_B,y_C,y_D]]\Theta_{[l_{Bx},h_{Bx},l_{By},h_{By}]}[x_{B+}[y_B,x_C,y_C,x_D,y_D],y_B]\\
&\Theta_{[0,\infty]}[D[y_B,x_C,y_C,x_D,y_D]]\\
&+\frac{F[x_{A-}[y_B,x_C,y_C,x_D,y_D],y_A[y_B,y_C,y_D],x_{B-}[y_B,x_C,y_C,x_D,y_D],y_B,x_C,y_C,x_D,y_D]}{\sqrt{D[y_B,x_C,y_C,x_D,y_D]}}\\
&\Theta_{[l_{Ax},h_{Ax},l_{Ay},h_{Ay}]}[x_{A-}[y_B,x_C,y_C,x_D,y_D],y_A[y_B,y_C,y_D]]\Theta_{[l_{Bx},h_{Bx},l_{By},h_{By}]}[x_{B-}[y_B,x_C,y_C,x_D,y_D],y_B]\\
&\Theta_{[0,\infty]}[D[y_B,x_C,y_C,x_D,y_D]]\Bigg)
\end{split}
\end{equation}
\end{widetext}

where $\Theta_{[...,...]}[...]$ represents the Heaviside function between the edges along the x and y directions (i.e $l_{Ax}$,$h_{Ax}$ or $l_{Ay}$,$h_{Ay}$ or $l_{Bx}$,$h_{Bx}$), of the elements corresponding to the variables reduced (in this case $x_A$ or $y_A$ or $x_B$), D is the discriminant of the quadratic equation obtained in the reduction of energy Dirac delta and the basis functions and $w^{e-e}_{0123}$, in Eq.~\eqref{ScatteringTensor}, are grouped in  F[...]. For a more detailed description of Eq.~\ref{MonteCarloIntegration} refer to Appendix \ref{Appendix:Phi}. 

Eq.\eqref{MonteCarloIntegration} does not include any Dirac deltas (yet it includes Heaviside step functions) and has been converted from a 8 dimensional integral (Eq.~\ref{ScatteringTensor}) to a 5 dimensional integral. It can be integrated with a number of techniques, and for simplicity we used standard Monte Carlo.

\section{Numerical tests for the code}\label{Sec:Tests}

We use as a numerical test case a simple one-band model. We choose one upper cone of graphene dispersion. We do not aim here at a very precise description of the thermalisation dynamics in graphene, nonetheless the considered band should be sufficient to adequately describe the thermalisation of doped graphene.

Moreover the used band dispersion is an excellent example to show the capability of the proposed method to describe with high precision the nature of the Dirac point of massless Dirac fermions.

We will include only electron-electron interaction. However, as already clarified, the method can be easily extended to different types of scatterings. We use as scattering amplitude, a constant. Notice that that is exactly correct for completely localised interaction, where the interaction is momentum independent, yet, of course, it is not correct in general. Moreover, it is important to appreciate that a very large component of the momentum and energy dependence of, for instance, lifetimes close to equilibrium comes from the phase space factor $P_{0123}$ in Eq.~\ref{ScatteringIntegral}, rather than the scattering amplitude itself. Nonetheless the full momenta dependence of the scattering amplitude (if known by other means that are not the object of this work) can be included in the numerical method.

We will use arbitrary units for momenta, scattering amplitude and time. 

\begin{figure}
\centering
\includegraphics[width=\columnwidth]{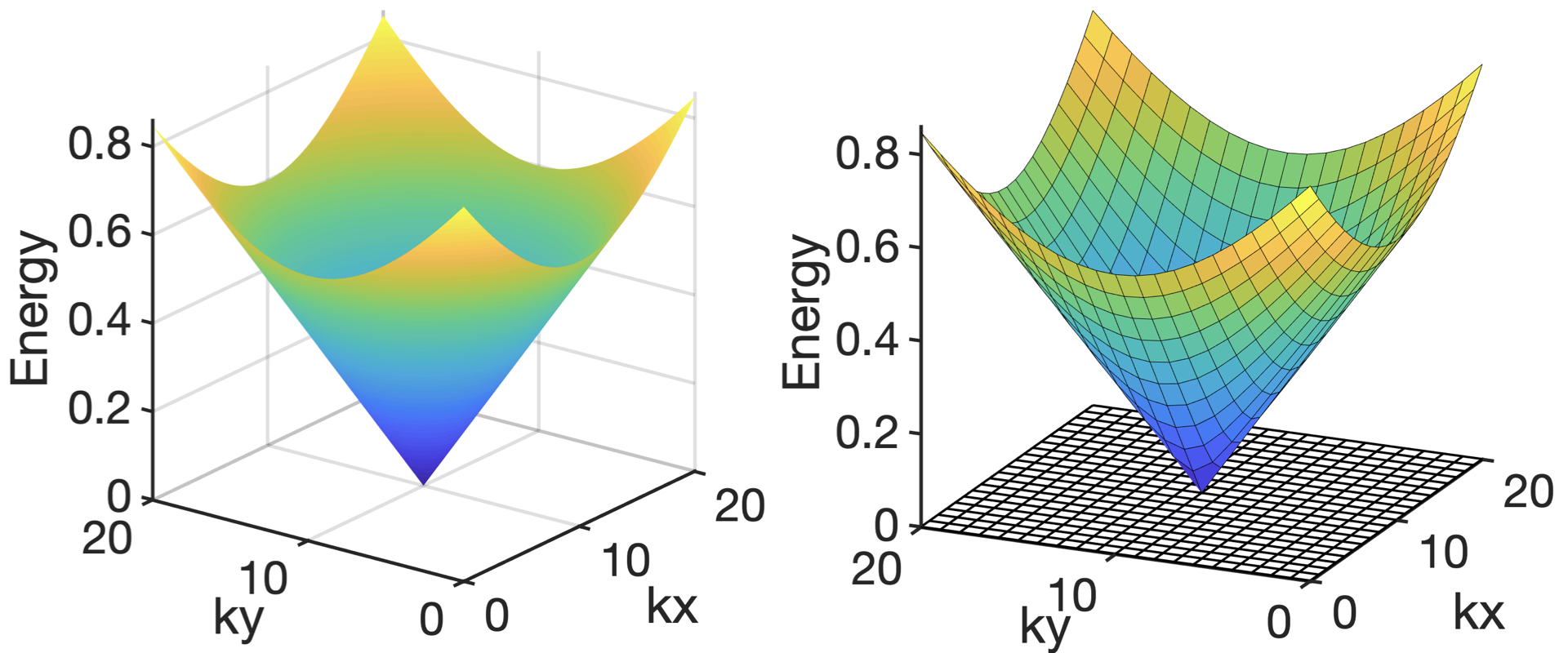}
\caption{Graphene band structure: Plotted from dispersion relation (left) and Representation generated by the numerical code (right) in the chosen basis functions. The mesh is a uniform rectangular grid with 20 elements in both, the kx and the ky axes.}
\label{fig:GrapheneBandStructure} 
\end{figure}

\subsection{Representation of the band structure}

 Fig.~\ref{fig:GrapheneBandStructure} shows how the exact dispersion of the Dirac cone is discretised using Eq.~\ref{fandEProjection}. Notice how the structure of the Dirac point is preserved. This is achieved only using the precaution of adjusting the mesh such that the Dirac point lies exactly on a mesh node. This is one of the advantages of using the chosen basis functions. 

\subsection{Scaling of computational cost and storage cost of the scattering tensor} \label{scaling}

\begin{figure}
\centering
\includegraphics[width=\columnwidth,height=6cm]{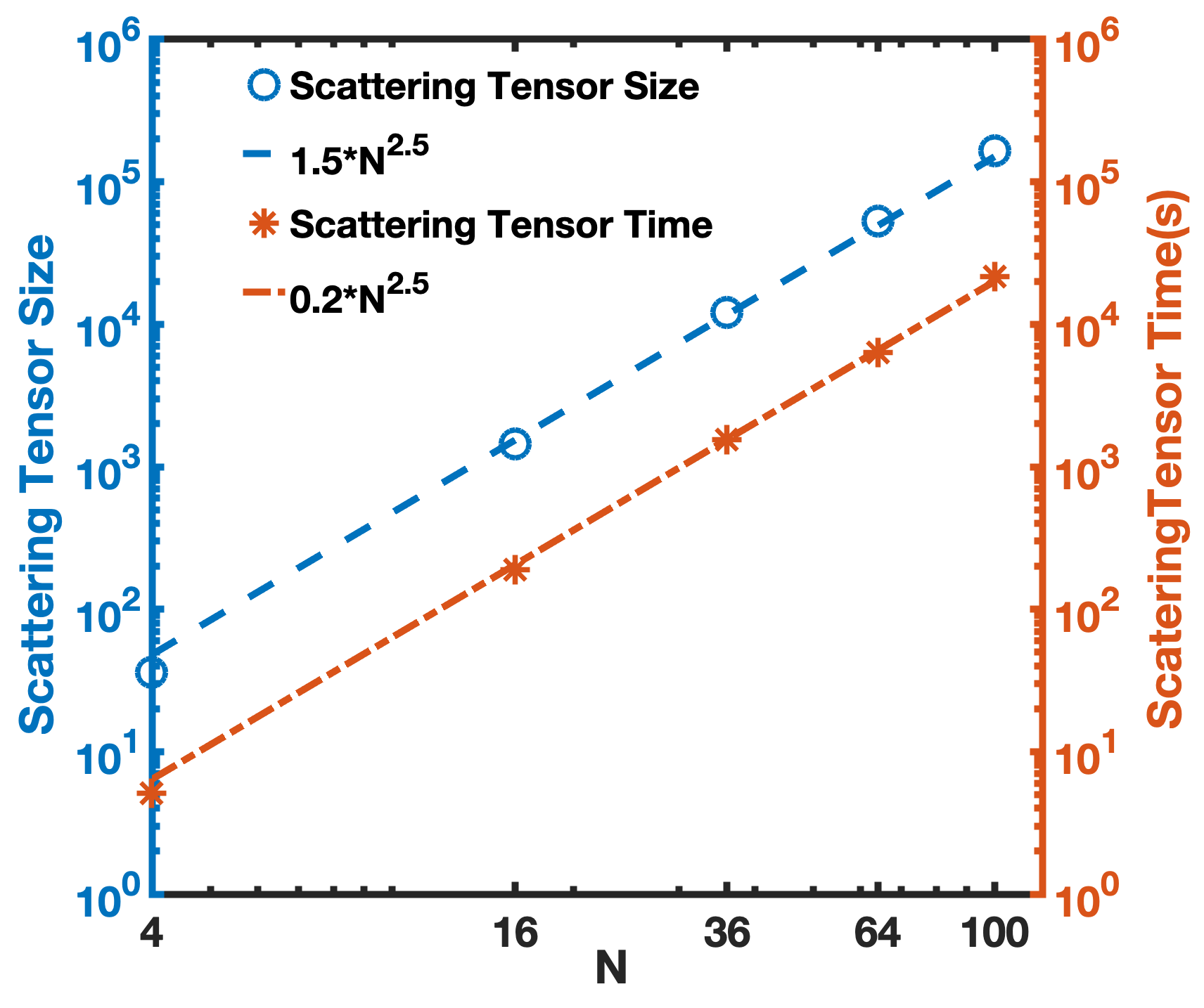}
\caption{Size of the scattering tensor and Computational time for the scattering tensor vs number of mesh elements}
\label{fig:Scal1} 
\end{figure}

In Ref.~\cite{Michael} we showed that with our numerical method, owing to our choice of basis functions, the scattering tensor size scales as $\sim N^{\frac{3d-1}{d}}$, where d is the dimensionality of the system and N is the total number of elements in the mesh (referred to as mesh resolution for convenience). Hence, for the 2D system considered in this study, the scattering tensor size should scale as $\sim N^{2.5}$. Similarly the actual wall time required for the calculation of the scattering tensor is another critical quantity of interest and we expect it to follow a similar scaling law. To verify this, in Fig.~\ref{fig:Scal1} we plot the number of entries in the scattering tensor and the computational time for the scattering tensor against an increasing mesh resolution, $N$. As seen from the figure, the size of the scattering tensor and the computational time indeed scale as $\sim N^{2.5}$. 

\subsection{Conservation of particles, momentum and Energy}

The finite stochastic error present in the Monte Carlo Integration method breaks the inherent symmetries of the scattering tensor which are equivalent to particle number, momentum and energy conservation. Failing to conserve these quantities could lead to spurious results and hence this is an extremely critical issue. However, when an apposite construction of the Monte Carlo points is done, these conservations are automatically obeyed\cite{Michael}. As such, we analyse the  particle number, momentum and energy conservation for our scattering tensor before actually using it for time propagation of the population. Indeed, the numerical code preserves these critical quantities to machine precisioncas seen from Table \ref{tab:Tab1}.

\begin{table}
    \centering
    \begin{tabularx}{\columnwidth} { 
  | >{\centering\arraybackslash}X 
  | >{\centering\arraybackslash}X 
  | >{\centering\arraybackslash}X | >{\centering\arraybackslash}X | >{\centering\arraybackslash}X | }
      \hline
        \textbf{Time Step} & \textbf{$\Delta P$} & \textbf{$(\Delta K)_x$} & \textbf{$(\Delta K)_y$} & \textbf{$\Delta E$}  \\
        \hline
         1& -1.38 \; e-14&-5.53 \; e-15&-1.19 \;e-13&-9.97 \;e-14\\
         \hline
         2&-1.27 \;e-14&-4.54 \;e-15&-1.02 \;e-13&-1.06 \;e-13\\
         \hline
         3&1.61 \;e-14&3.98 \;e-15&1.25 \;e-13&1.21 \;e-13\\
         \hline
         4&-3.41 \;e-15&-1.10 \;e-15&-2.07 \;e-14&-2.09 \;e-14\\
         \hline
    \end{tabularx}
    \caption{Change in particle number ($\Delta P $), Change in momentum along X direction ($(\Delta K)_x$), Change in momentum along Y direction ($(\Delta K)_y$) and Change in the total energy ($\Delta E$) over 4 time steps.}
    \label{tab:Tab1}
\end{table}

\subsection{Discretisation of the Fermi Dirac distribution}

The Fermi Dirac distribution is a steady state of the scattering operator in Eq.~\ref{ScatteringIntegral}. However the discretised Fermi Dirac distribution is not necessarily a steady state for the discretised scattering operator in Eq.~\ref{CollisionIntegral}. This discrepancy can be exacerbated at low resolution. 

For that reason, before starting the simulation, we test if the resolution is sufficient to resolve both the Fermi Dirac (at a desired temperature) and the thermalisation process. After discretising the dispersion we initialise the population to the discretised Fermi-Dirac population, then we let the time propagation proceed and note how far the discretised Fermi Dirac is from the actual numerical thermal equilibrium.

We initialise the population as a Fermi Dirac distribution with \(\mu=0.1eV\)  and T=700 K calculated on the band structure depicted in Fig.~\ref{fig:GrapheneBandStructure}. To use this mentioned function as a numerical steady state solution we want to evaluate how close it is to the numerical thermal equilibrium and so we propagate the population in time for 300 time steps with a time step of $dt=0.0001$ (below, we will see that typical lifetimes are of the order of $10^{-3}$) . 

\begin{figure}
\centering
\includegraphics[width=\columnwidth,height=8cm]{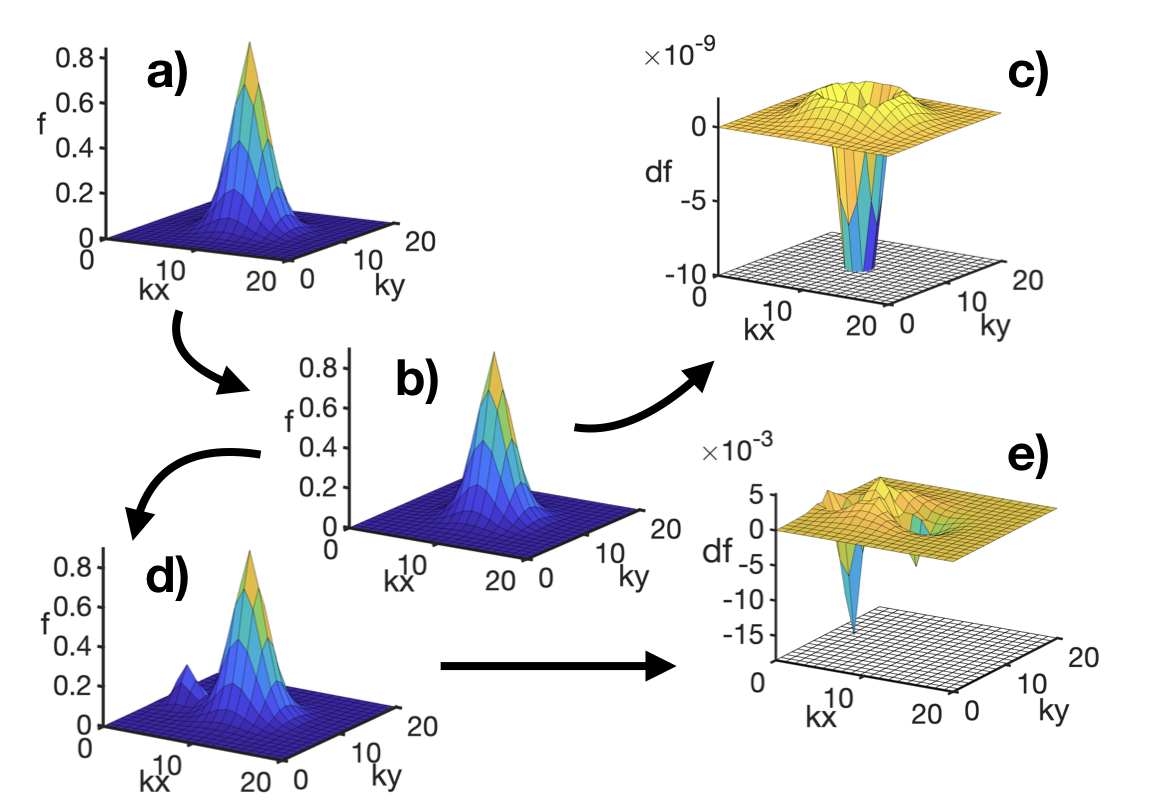}
\caption{Check for confirming that Fermi Dirac distribution is the steady state solution. a) Initial input Fermi-Dirac distribution b) Numerical Steady State solution c) Change in the population from the numerical steady state solution after 1 time step of dt=1e-4 d) Small excitation added to the numerical steady state solution e) Change in the population from the numerical steady state solution post addition of the excitation after 1 time step of dt=1e-4. Note the orders of magnitude change in the population post addition of excitation as compared to the change in the population after reaching the numerical steady state solution.}
\label{fig:SteadyState} 
\end{figure}

Then, to estimate the error of the new improved numerical thermal equilibrium, we propagate  the distribution for one more time step and note the change in population. Now we introduce a small number of particles in the system, which corresponds to an excitation, and note the change in the population after 1 time step (see Fig.~\ref{fig:SteadyState}). Given that the first derivative is negligible compared to the derivative in the second case (which is more representative of the dynamics we are interested in studying) it can be safely concluded that the numerically obtained distribution after 300 time steps is sufficiently close to the steady state solution. 

\section{Results and Discussion}

\begin{figure}
\centering
\includegraphics[width=\columnwidth]{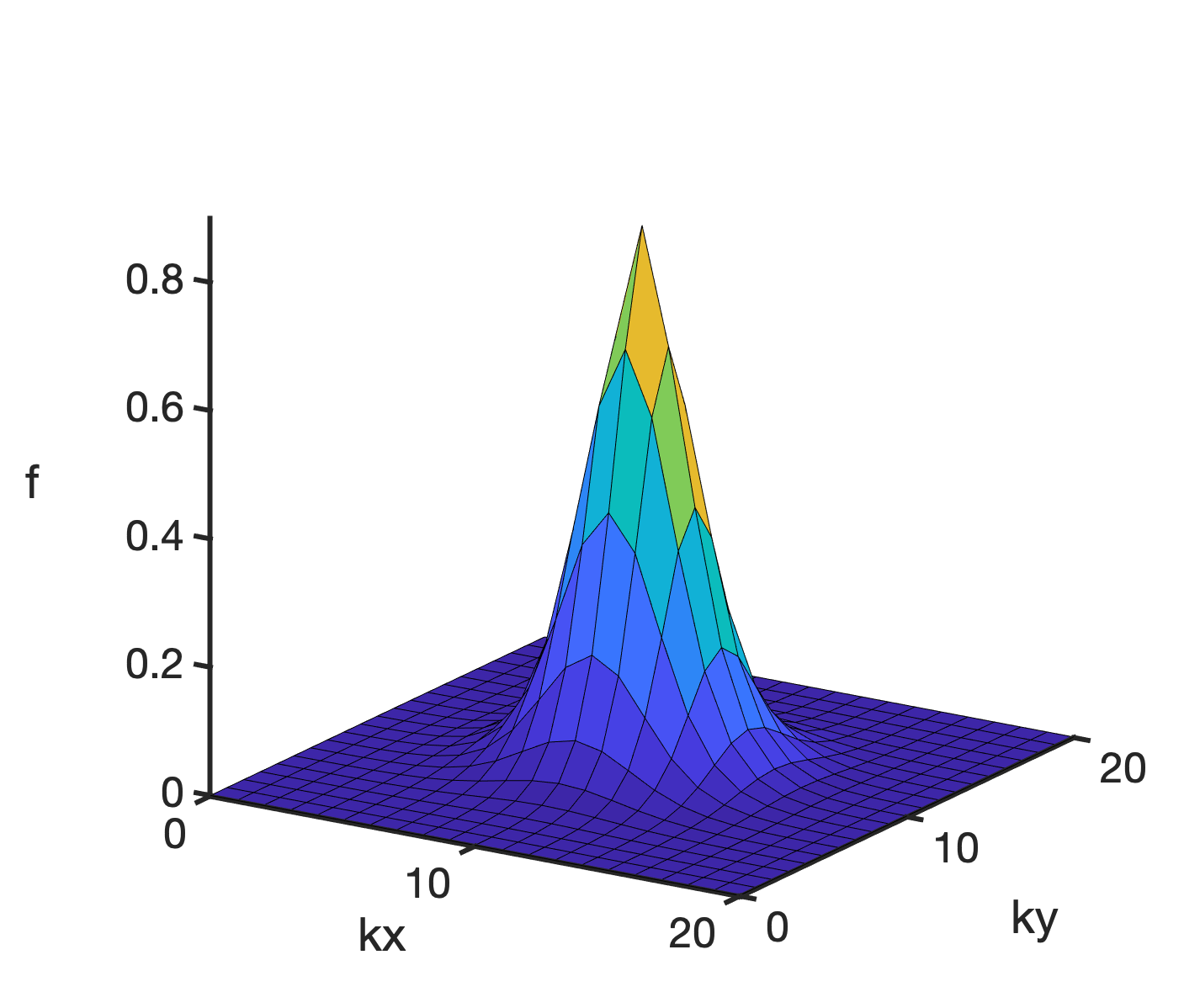}
\caption{Numerical steady state distribution after stabilizing the input Fermi-Dirac distribution for 300 timesteps.}
\label{fig:InpFD} 
\end{figure}

In this section we apply the code to a selected 2D system and study the thermalisations of the added out-of-equilibrium excitations. Since it is not our intention to study real systems, we only demonstrate the capabilities of the code considering excitations in doped graphene with a bandstructure shown in Fig.~\ref{fig:GrapheneBandStructure}.  We follow the methodology below:
\begin{enumerate}
    \item \textbf{Choose a mesh resolution:} In this study the mesh resolution was kept to 20 elements in the x and y directions each.
    \item \textbf{Calculate the list of element combinations with possible scatterings:} We traverse through all the possible combinations of elements, corresponding to the scattering channel for which we wish to calculate the scattering tensor, to find a list of combinations where momentum and energy conservation can be satisfied.
    \item \textbf{Calculate the scattering tensor using this list:} The list obtained in the previous step is used to calculate the scattering tensor corresponding to the scattering channel considered. This scattering tensor will be used to time propagate the excitations and study the thermalisation characteristics.
    \item \textbf{Introduce an initial Fermi Dirac and let it  stabilise to an equilibrium distribution as described in section \ref{Sec:Tests}.} The initial Fermi Dirac distribution is introduced at a chemical potential, \(\mu=0.1eV\)  and Temperature, \(T=700 K\). We then let it stabilise to a numerical steady state distribution as described in section \ref{Sec:Tests}. Now we have a numerical steady state solution as shown in Fig.~\ref{fig:InpFD}.
    \item \textbf{Introduce an excitation in this numerical steady state solution and propagate in time}. The introduced excitation is propagated in time using DP853 time propagation scheme. We monitor the change in population as it decays to a new equilibrium distribution and analyze the thermalization dynamics.
\end{enumerate}

Initially we study small excitations over the equilibrium Fermi-Dirac distribution before moving on to excitations which resemble the ones in real experiments of ultrafast optical laser excitations of doped graphene.


\subsection{Scattering Rates}
\begin{figure}
\centering
\includegraphics[width=\columnwidth]{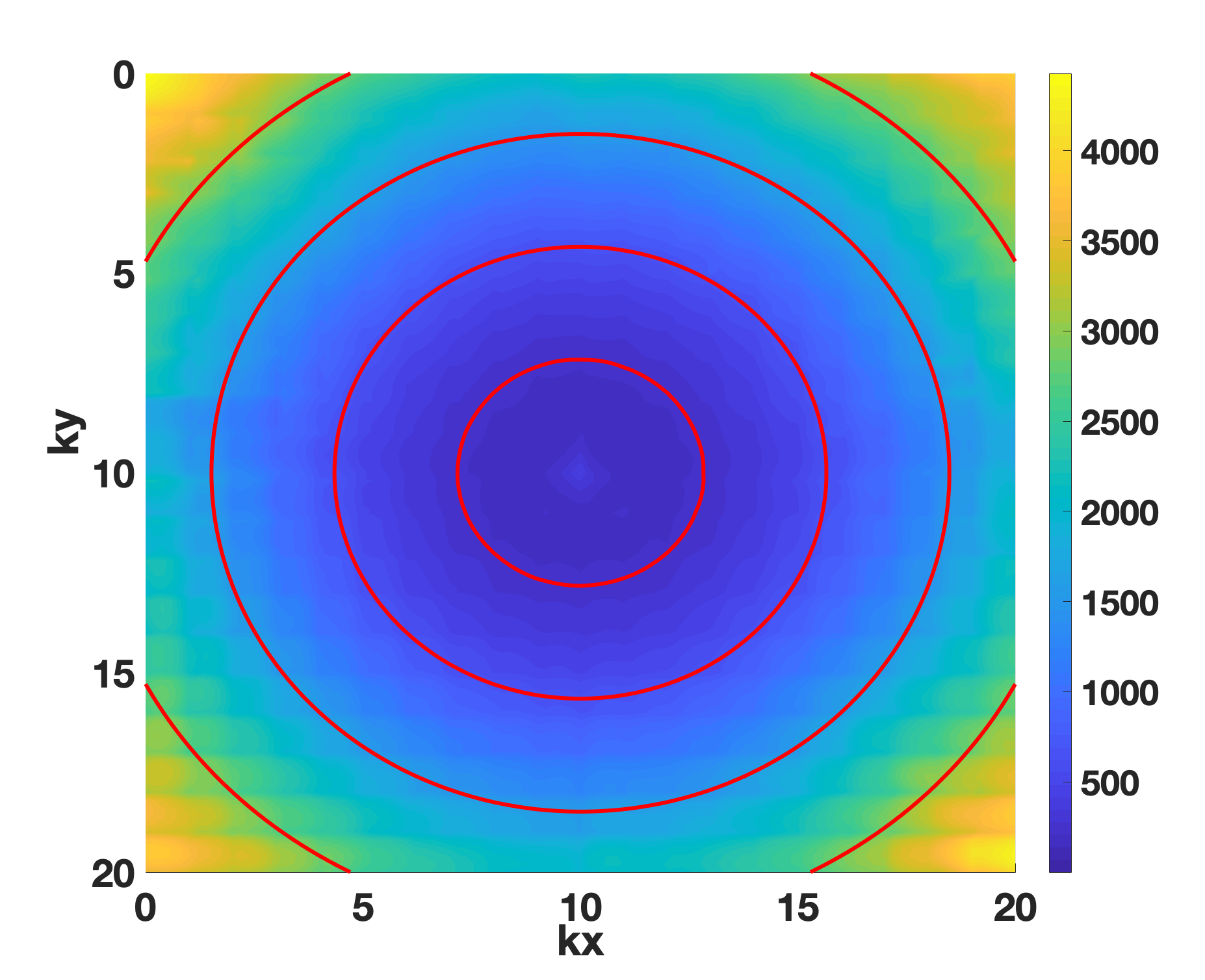}
\caption{Equilibrium Scattering Rates for an electron in the band depicted in fig.\ref{fig:GrapheneBandStructure}. Equal energy contours are also shown in red. Higher energies correspond to higher scattering rates.}
\label{fig:ScatRate} 
\end{figure}

Before studying the decay of excitations, we analyze the scattering rates for electron-electron interaction for the band structure depicted in Fig.~\ref{fig:GrapheneBandStructure} (see Appendix \ref{Appendix:ScatRates} or Ref.~\cite{Michael}). Fig.~\ref{fig:ScatRate} shows the scattering rates juxtaposed with equal energy lines in red. Particles added at higher energies should decay faster thereby giving higher scattering rates. For the inverted cone band structure shown in Fig.~\ref{fig:GrapheneBandStructure}, energy increases radially from the center of the Brillouin zone. Accordingly, Fig.~\ref{fig:ScatRate} depicts radially increasing scattering rates as expected.

\subsection{Decay of momentum-localised excitations}

\begin{figure}
\centering
\includegraphics[width=\columnwidth]{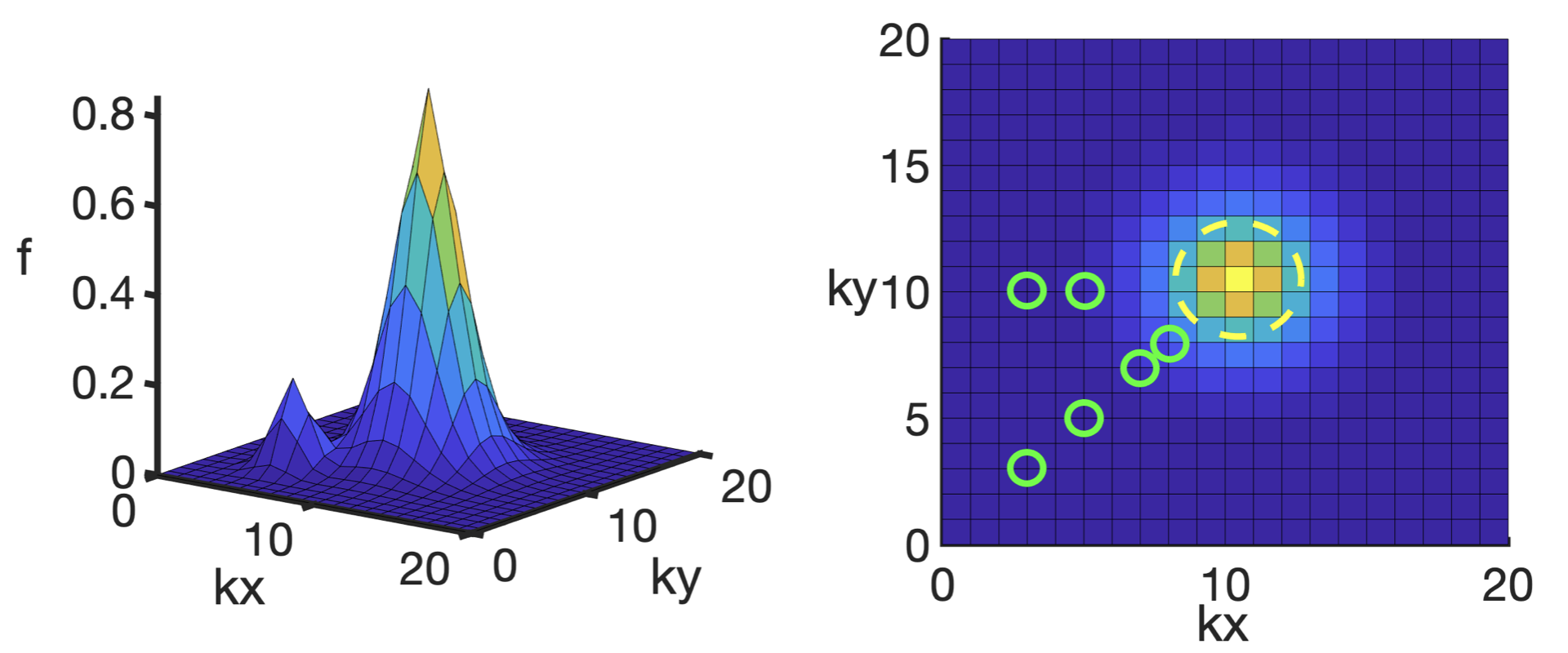}
\caption{Introduced excitations with the initial FermiDirac distribution (left) and Locations of the introduced excitations in the Brillouin zone given by full circles (right). The dotted circle at the center highlights the location of the initial Fermi-Dirac distribution.}
\label{fig:BumpLoc} 
\end{figure} 

As a first step we run a few test cases with excitations that correspond to the introduction of particles in a small region of the Brillouin zone over the numerical steady state distribution in Fig.~\ref{fig:InpFD}. Given that the band structure and the initial numerical steady state distribution are both radially symmetric, it suffices to study different excitations only in one quadrant along the diagonal and along a line perpendicular to one side of the domain. Fig.~\ref{fig:BumpLoc} shows the locations where excitations are introduced, while the exact coordinates on the mesh for the introduced excitations are listed in table \ref{tab:Tab2}. Notice that, even if excitations in the two directions should be equivalent, due to the shape of the considered part of the Brillouin zone, they are not exactly equivalent anymore.

\begin{figure}
\centering
\includegraphics[width=\columnwidth]{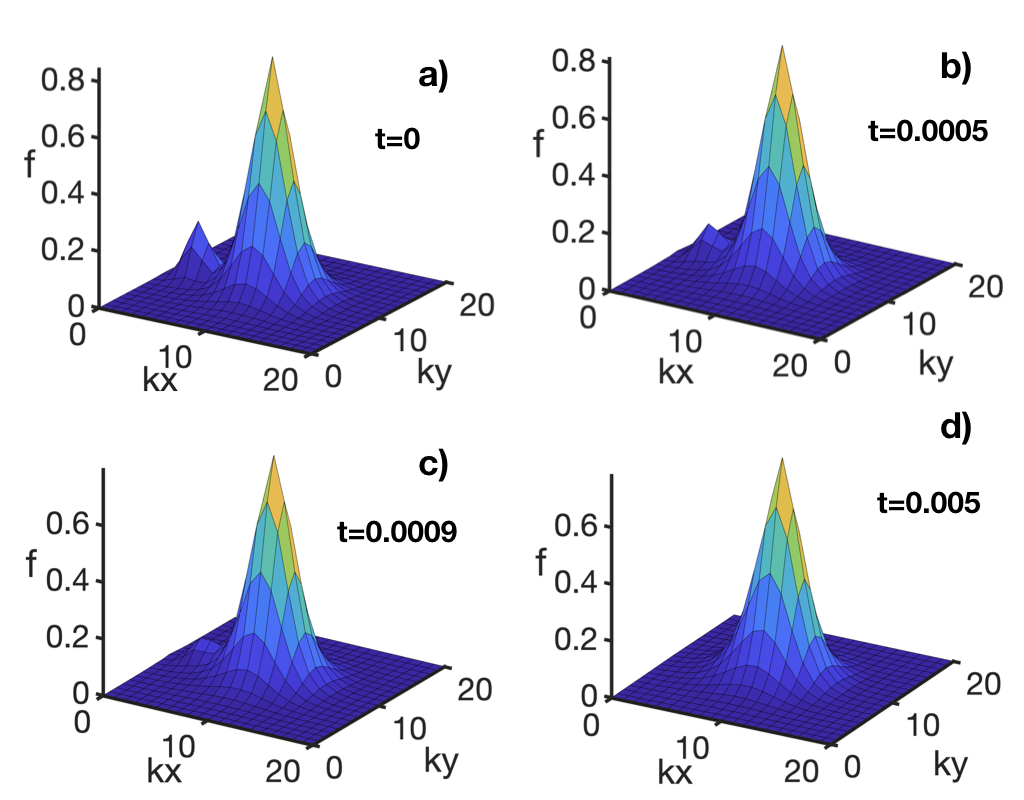}
\caption{Decay of an introduced excitation with time. We show here the change in the population distribution function only for the excitation at (3,10).}
\label{fig:DecayBump} 
\end{figure}

\begin{figure}
\centering
\includegraphics[width=\columnwidth]{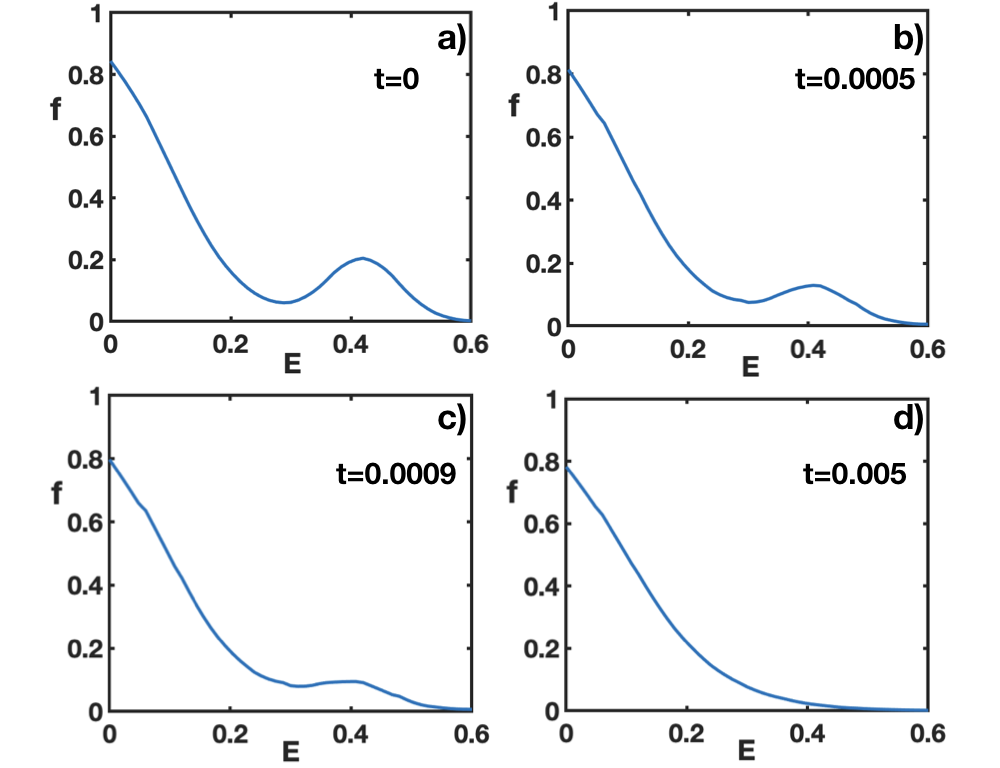}
\caption{Variation of population distribution function for the excitation at (3,10) with respect to energy as the thermalisation progresses.}
\label{fig:DecayfvsEBump} 
\end{figure}

Figure \ref{fig:DecayBump} shows the decay of one of the excitations at the momentum  locations depicted on the right in Fig.~\ref{fig:BumpLoc}. As the thermalization progresses, the population distribution function varies smoothly and relaxes to a new equilibrium distribution. Fig.~\ref{fig:DecayfvsEBump} shows the change of the population distribution function with respect to energy over the [(10,10), (0,10)] line of the Brillouin zone (passing through the introduced excitation), as the thermalization progresses. 

Since we do not consider Umklapp scatterings here, the initial momentum cannot be dissipated.  The main implication is that the steady state distribution cannot be a Fermi-Dirac distribution. Nevertheless, from Fig.~\ref{fig:DecayfvsEBump}, we extract two quantities of interest as below:
\begin{align}
        \beta =&-\frac{\partial f}{\partial E} \Bigg|_{f=0.5} \\
        \mu=&E|_{f=0.5}
\end{align}

The first term, $\beta$, is analogous to the $\frac{1}{K_B T }$ term in the Fermi-Dirac equilibrium distribution and we use it here as an indicator of the temperature. The second term, $\mu$, is analogous to the chemical potential term in the Fermi-Dirac equilibrium distribution. Together, the variation of $\beta$ and $\mu$ can provide interesting insights into how the population distribution function changes its shape.

\begin{figure}
\centering
\includegraphics[width=\columnwidth]{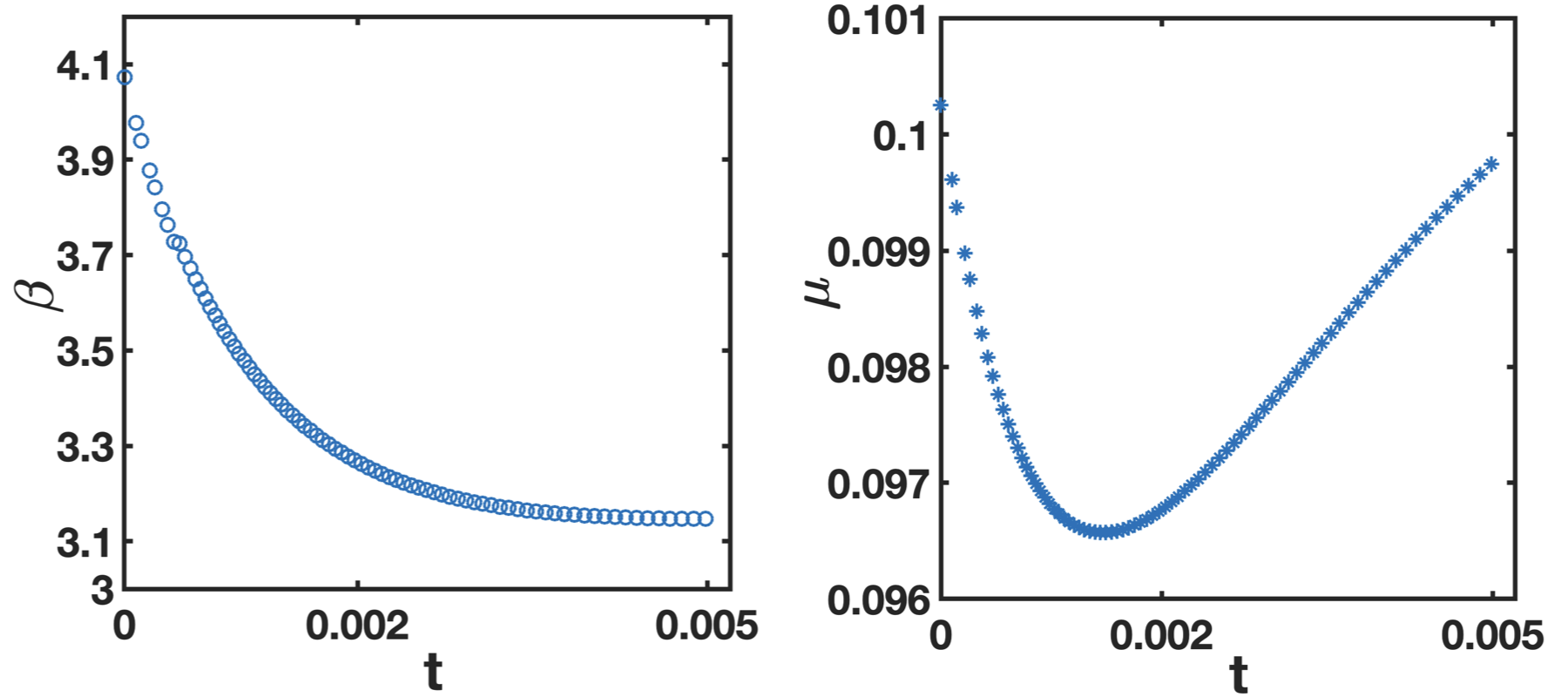}
\caption{Variation of $\beta$ and $\mu$ for the excitation at (3,10) with time.} 
\label{fig:MuBetaBump} 
\end{figure}

Fig.~\ref{fig:MuBetaBump} shows the variation of $\beta$ and $\mu$ with time for the same excitation considered in Fig.~\ref{fig:DecayBump}. At time $t=0$, the value of $\beta$ corresponds to that of the numerical steady state distribution, which is a Fermi-Dirac distribution, and then as the thermalization progresses $\beta$ decreases continuously indicating an increase in the overall temperature. This was expected since the added excitation was at higher energies and energy conservation dictates that the thermalised distribution should therefore be at higher energy or higher temperature than the numerical steady state distribution. The variation of $\mu$ is, on the other hand, very small.

\begin{figure}
\centering
\includegraphics[width=\columnwidth]{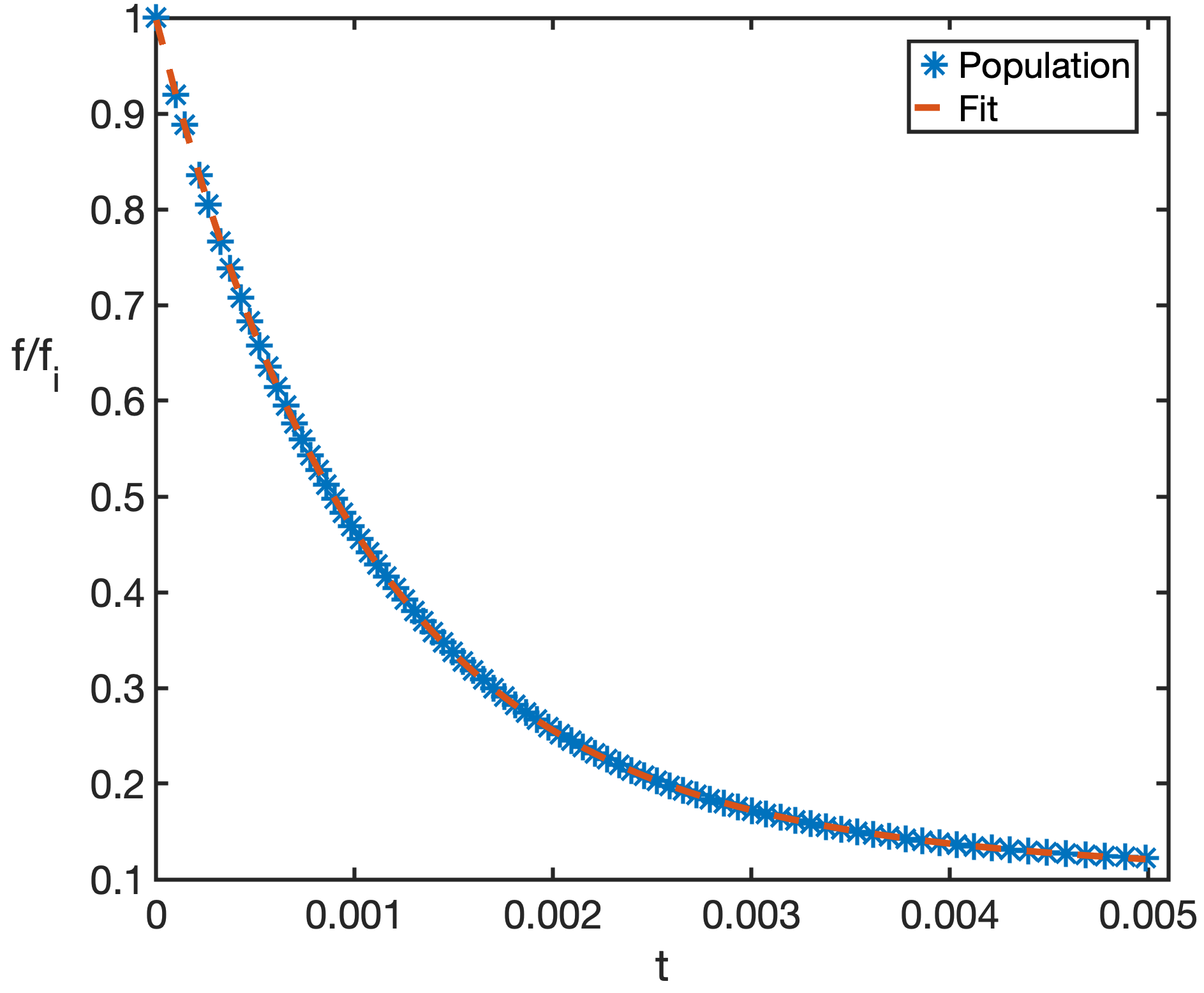}
\caption{Decay at the peak of the introduced excitation from Fig.~\ref{fig:DecayBump} with time. For comparison we also show a double exponential fit which has the expression as: $0.8573*exp(-t/\tau)+ 0.1405*exp(-t/\tau')$ with $\tau=0.001$ and $\tau'=0.024$.}
\label{fig:DecayFitBump} 
\end{figure}

Fig.~\ref{fig:DecayFitBump} shows the time dependence of the population at the point in the Brillouin zone where the introduced excitation is localised. We normalize the values of the population distribution function at this point by its initial value. The population decay is a non-trivial function of time and not a simple exponential, which is expected only for small excitations. However, for comparison we show a double exponential fit. Although a double exponential fit is still not sufficient and does not completely represent the thermalisation as seen from figure \ref{fig:DecayFitBump}, nonetheless, it can still help to generate a quantitative estimation of the time scales involved in the thermalisation. $\tau$ represents the initial time scale of decay, which is generally smaller, while $\tau'$ represents the thermalisation time scale at later stages, which is generally longer. 

\begin{table}
    \centering
    \begin{tabularx}{\columnwidth} { 
  | >{\centering\arraybackslash}X 
  | >{\centering\arraybackslash}X 
  | >{\centering\arraybackslash}X 
   | }
      \hline
        Excitation No. & Co-ordinates of the Excitation on mesh (X,Y) & Inverse decay times $(1/\tau)$\\
         \hline
        1 & (8,8) & 221.7 \\
        \hline
        2 & (7,7) & 417.9 \\
        \hline
        3 & (5,10) & 528.1 \\
        \hline
        4 &(3,10) & 822.6 \\
        \hline
        5 &(5,5) & 1347 \\
        \hline
        6 &(3,3) & 2558\\
        \hline
    \end{tabularx}
    \caption{Coordinates of the introduced excitations in this study. The distance from the center of the Brillouin zone and hence the energy of the introduced excitation increases from excitation no.1 to no.6 progressively. Using a double exponential fit as in fig.\ref{fig:DecayFitBump} we compare the initial scattering rates ($1/\tau$) for each excitation. Excitations at higher energies are expected to decay faster and accordingly the scattering rates increase from excitation no.1 to no.6.}
    \label{tab:Tab2}
\end{table}

\begin{figure}
\centering
\includegraphics[width=\columnwidth]{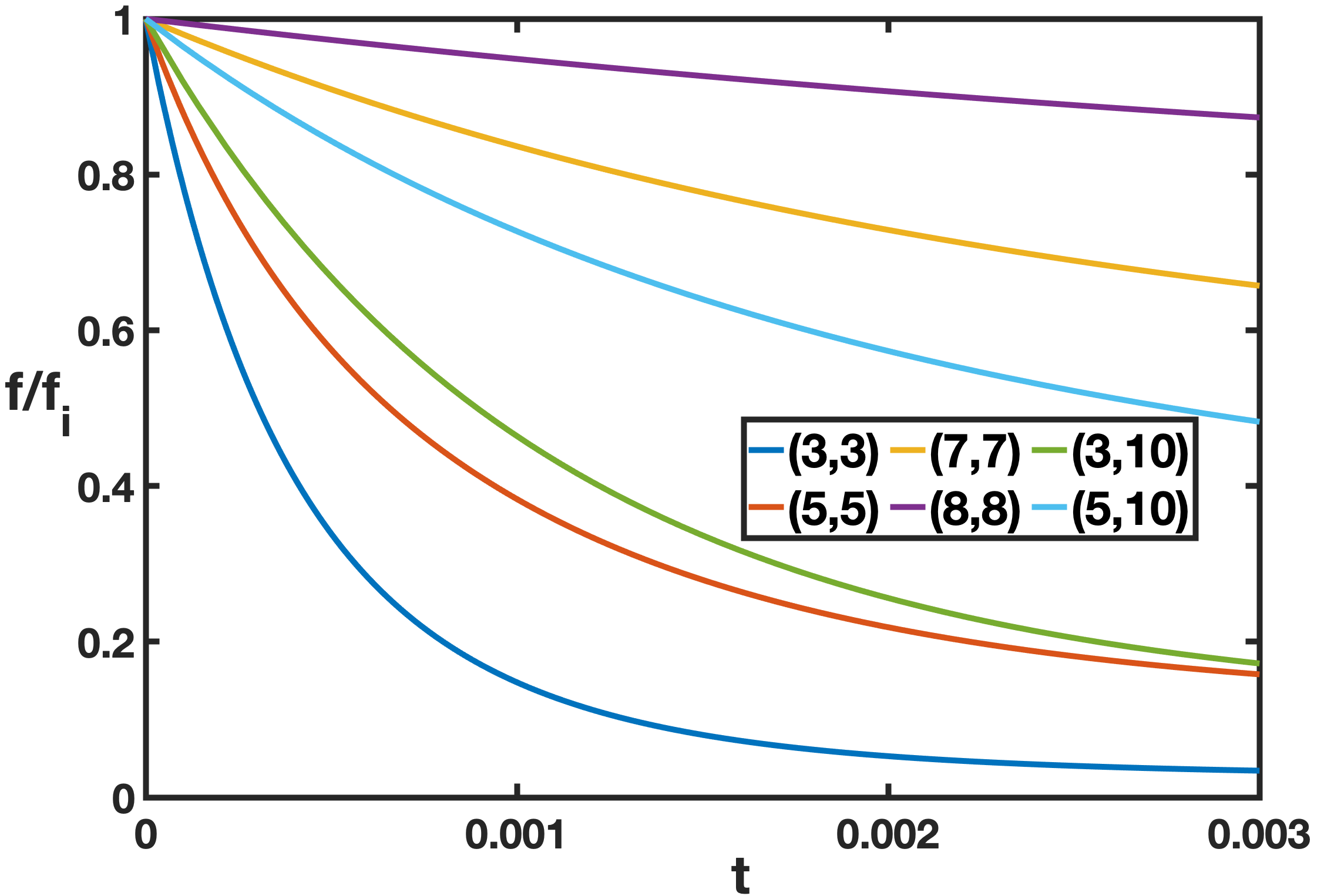}
\caption{ Thermalization of introduced excitations with time. For a direct comparison, we plot the normalised population distribution function at the peak of the introduced excitation vs time. The initial value of the population distribution function, $f_i$, is used for the normalisation. Excitations introduced further away from the center of Brillouin zone are at higher energies and hence they thermalise relatively quickly.}
\label{fig:GraphDecayBumps} 
\end{figure}

We repeat the same analysis for same excitations localised at all the different momentum regions in Fig.~\ref{fig:BumpLoc}.b, and listed in table \ref{tab:Tab2}. To compare the thermalisation of all the introduced excitations on a common footing we again normalize the values of the population distribution function at the central momentum coordinates of each excitation by its initial value and compare the normalized thermalisation profiles in Fig.~\ref{fig:GraphDecayBumps}. The excitations away from the center (which are at higher energies) are expected to decay faster. Indeed, Fig.~\ref{fig:GraphDecayBumps} shows that the decay speeds up as the excitations move away from the center. Notice that, again the excitation cannot be considered small, and therefore the decay is not exponential, neither decays with a time constant (the lifetime) which is the inverse scattering rate. Nonetheless, from Table \ref{tab:Tab2}, we observe that the decay constants far from equilibrium remain similar to the inverse scattering rates in Fig.~\ref{fig:ScatRate}, calculated at equilibrium.

\subsection{Decay of realistic excitations}

\begin{figure}[tb]
\centering
\includegraphics[width=\columnwidth]{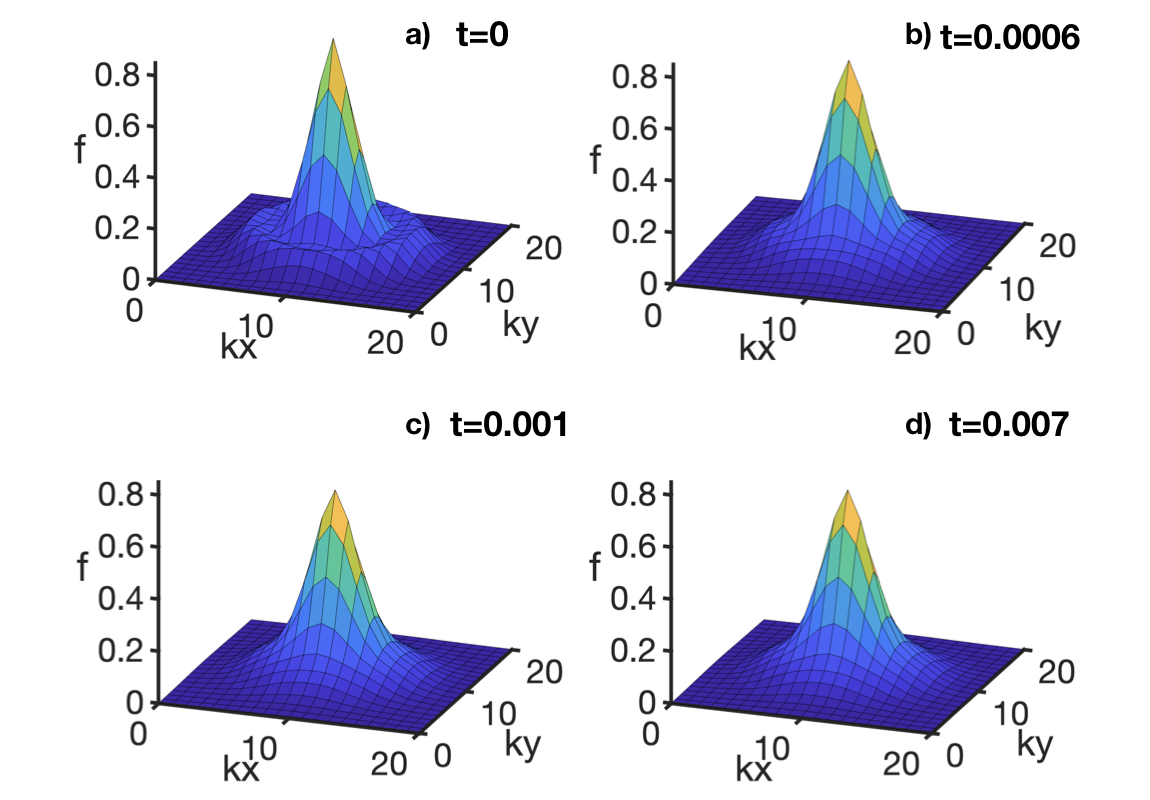}
\caption{ Thermalization of realistic small excitation, at a radius of 6 (1/nm), with time. The total population distribution with the introduced gaussian excitation thermalises to a new equilibrium distribution with time. }
\label{fig:Exc1} 
\end{figure}

It is not our aim here to provide a very realistic description of femtosecond optical excitations in graphene, yet we here highlight how the thermalisation after more realistic excitations can be addressed. In reality, for doped graphene, when light excites electrons from the valence band to the conduction band, the excitation will not manifest as a momentum-localised one, as analysed in the last section. A laser will excite transitions with an energy close to a central one. If we assume that the main transitions are from the valence Dirac cone to the conduction one, a better approximation for the initial population after the excitation would be the one in Fig.~\ref{fig:Exc1}.a. Notice how excited particles around a given energy have been added to the equilibrium distribution.

\begin{figure}[tb]
\centering
\includegraphics[width=\columnwidth]{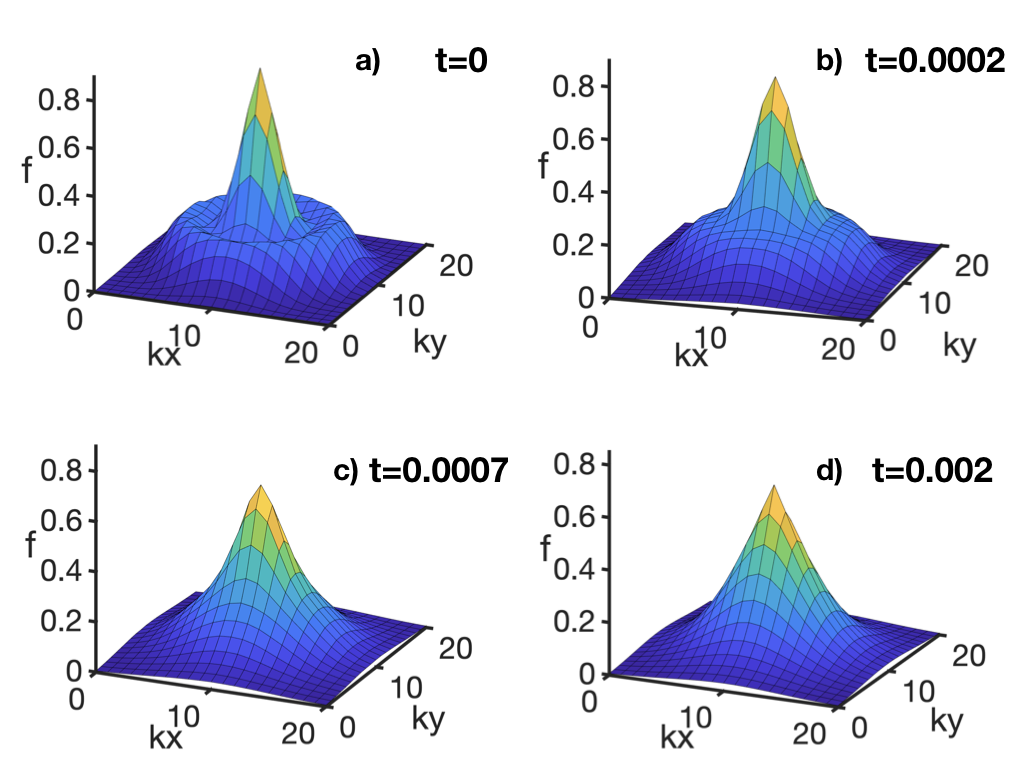}
\caption{ Thermalization of realistic strong excitation with time}
\label{fig:Exc} 
\end{figure}

\begin{figure}
\centering
\includegraphics[width=\columnwidth]{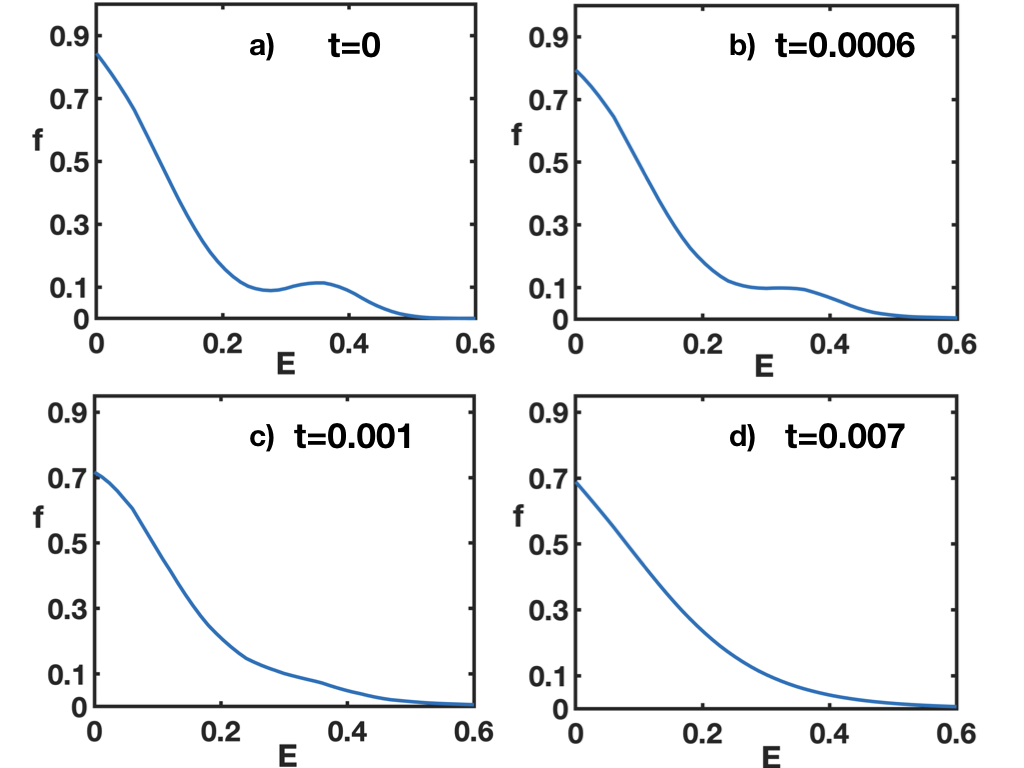}
\caption{ Energy resolved change in the distribution function of the introduced small excitation in fig.\ref{fig:Exc1} with time}
\label{fig:fVsESmall} 
\end{figure}

To analyze the thermalization of such cases we simulated 2 cases: one with a small excitation, introduced at a momentum radius of 6 (1/nm) from the center of the Brillouin zone, and another with a heavy excitation at a momentum radius of 6 (1/nm). For both cases the population distribution function varies smoothly to a new equilibrium distribution as seen from Fig.~\ref{fig:Exc1} and Fig.~\ref{fig:Exc} respectively. Similarly to the case of momentum-localised excitations considered previously, we can analyse the variation of the population distribution function with energy along any section of the Brillouin zone (the excitation is radially symmetric) as shown in fig.\ref{fig:fVsESmall}. 

\begin{figure}
\centering
\includegraphics[width=\columnwidth]{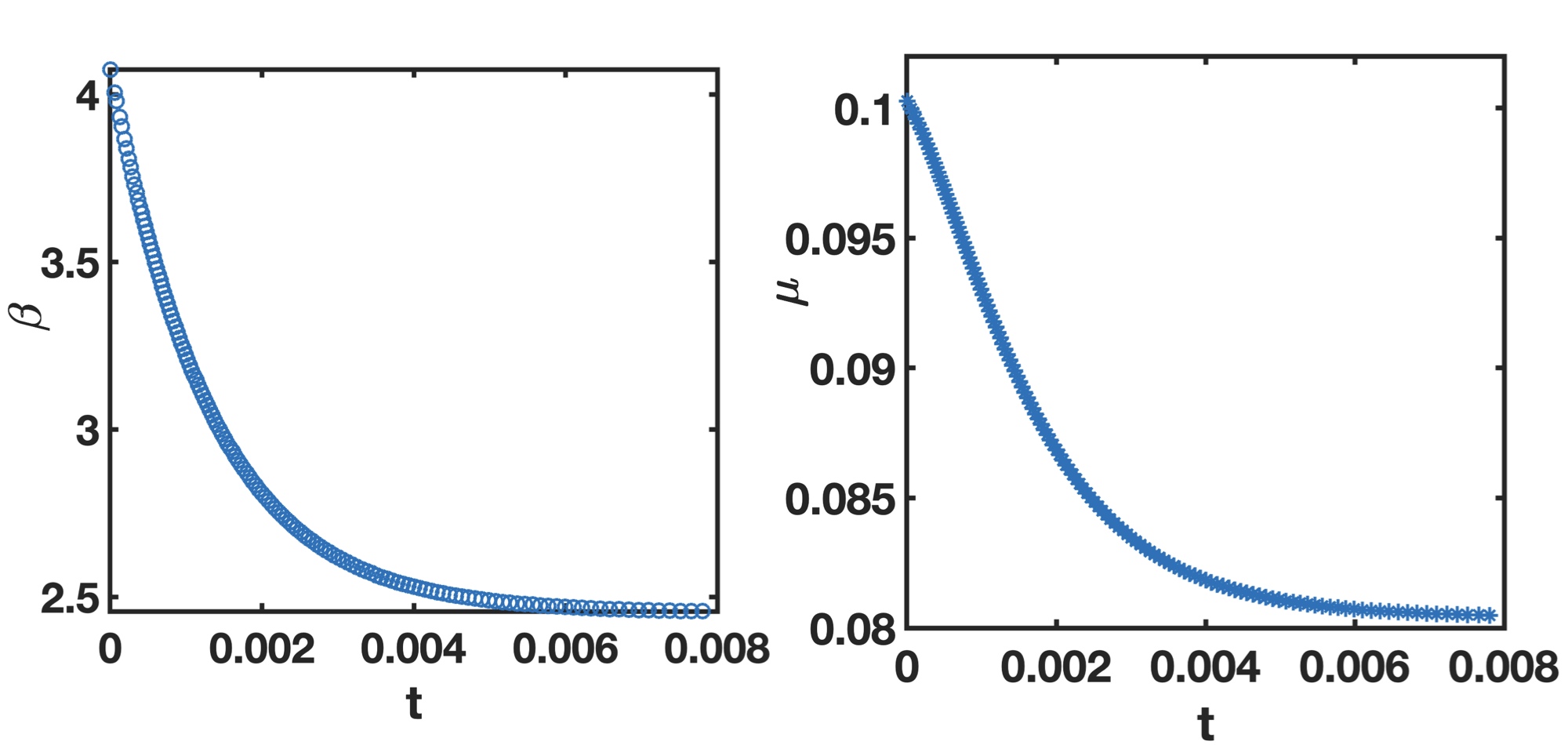}
\caption{ Variation of $\beta$ and $\mu$ with time for the introduced small gaussian excitation.}
\label{fig:BetaMuSmall} 
\end{figure}

\begin{figure}
\centering
\includegraphics[width=\columnwidth]{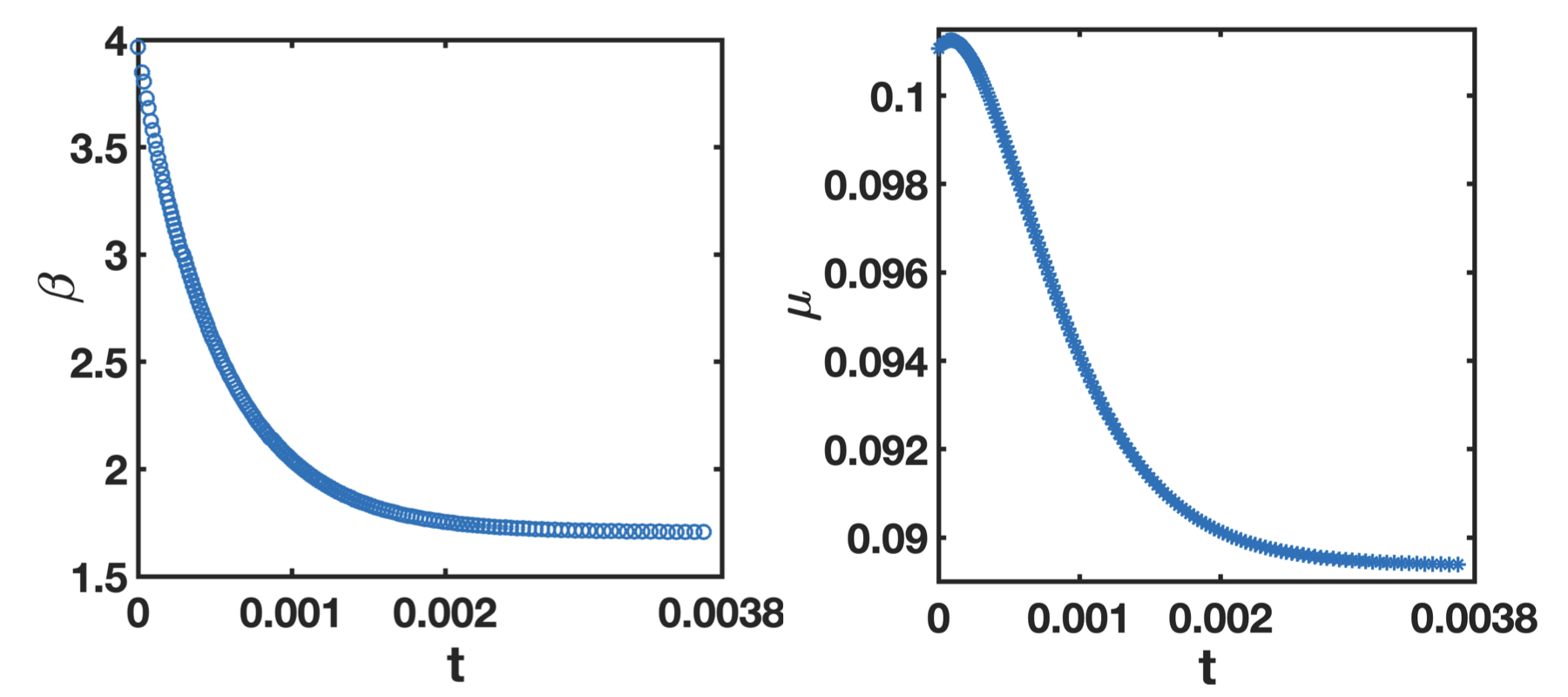}
\caption{Variation of $\beta$ and $\mu$ with time for realistic heavy excitation.}
\label{fig:BetaMuBig} 
\end{figure}

\begin{figure}
\centering
\includegraphics[width=\columnwidth]{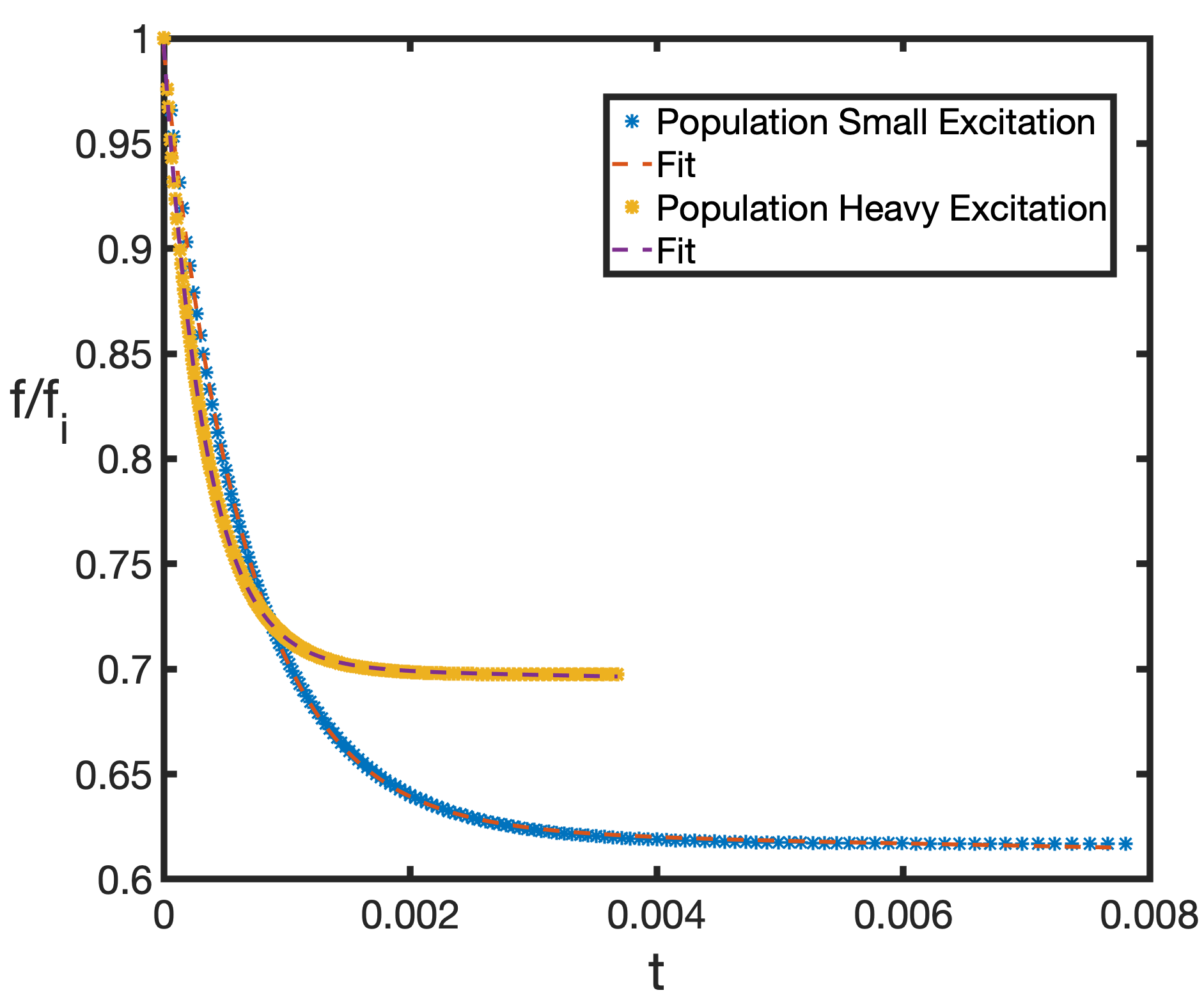}
\caption{Decay at the peak of the introduced a) small and b) realistic heavy excitations with time. A double exponential fit is also shown for comparison. The decay constant can be inferred from the time constants, $\tau$ and $\tau'$ in the exponential fit for each decay profile. The double exponential fit used for the decay of the small excitation is: $0.371*exp(-t/\tau)+ 0.6238*exp(-t/\tau')$ with $\tau$=0.00065 and $\tau'$=0.5485, while the one used for decay of the realistic heavy excitation is: $0.2965*exp(-t/\tau)+ 0.7007*exp(-t/\tau')$ with $\tau$=0.00033 and $\tau'$=0.6045.} 
\label{fig:DecayFitExcitation} 
\end{figure}

The variation of $\beta$ and $\mu$ is shown in Fig.~\ref{fig:BetaMuSmall} and \ref{fig:BetaMuBig} respectively. As in the previous case of isolated excitations, beta decreases monotonically indicating an increase in temperature. The variation of $\mu$ is now much larger than in the previous case. This is to be expected given that the number of particles added in the band by the present excitation is much larger. Close to equilibrium, the Fermi level usually changes with changing temperature to prevent the change of the total number of particles. In this case, the dynamic estimation of the Fermi level has a similar role. Given the large energy the added particles had right after the excitation, the dynamic temperature grew considerably, pushing the chemical potential to lower values. In Fig.~\ref{fig:BetaMuSmall} it is also possible to observe how, higher energy excitations lead to faster thermalisation dynamics compared to lower energy ones (as one would expect close to equilibrium when observing scattering rates). Fig. \ref{fig:DecayFitExcitation} also highlights that larger excitations added at the same energy thermalise faster.

Finally, in Fig.~\ref{fig:DecayFitExcitation}  we present the decay with time at the peak of the introduced gaussian excitations. As before, we normalise the population distribution function at the central momentum coordinates with its initial value. For comparison we again fit a double exponential function and generate a quantitative estimate of the time scales involved in the thermalisation. We remind that the value of the scattering amplitude, $w^{e-e}_{0123}$ was chosen to be a constant equal to 1 in these simulations. Fitting the decay of the introduced excitation from our simulations (eg.~Fig.~\ref{fig:DecayFitExcitation}) to an actual decay data from experiments, we can conclude on an appropriate value of the scattering amplitude, $w^{e-e}_{0123}$ thereby fixing the only free parameter in our method.

\begin{figure}
\centering
\includegraphics[width=\columnwidth]{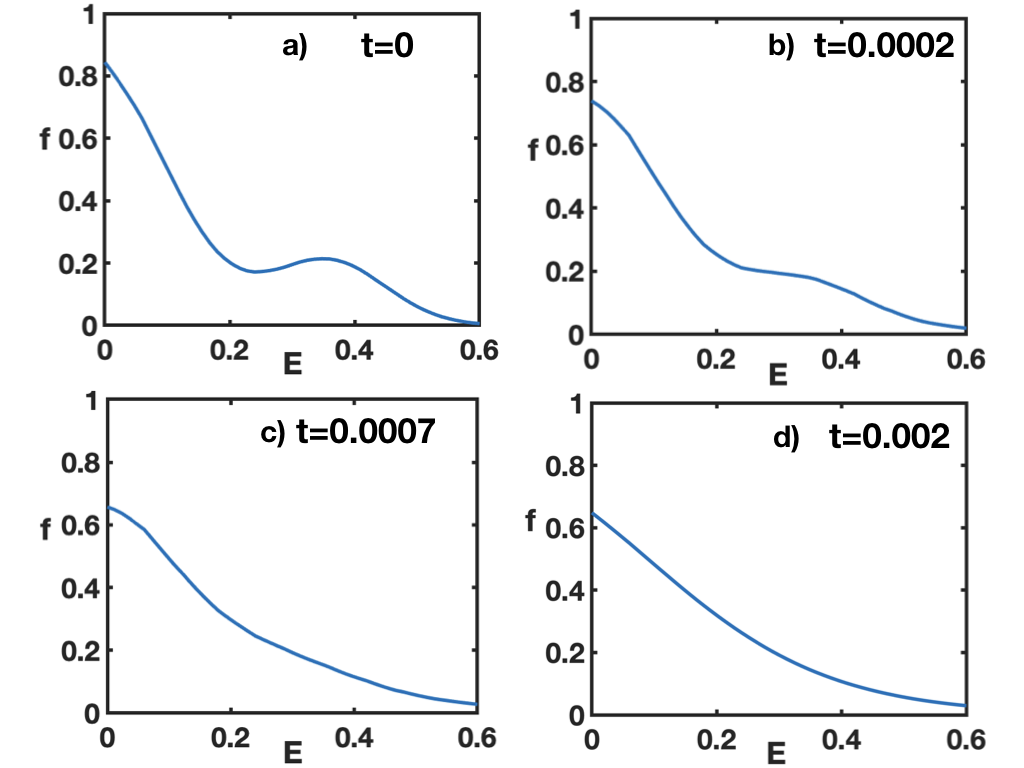}
\caption{ Energy resolved change in the distribution function of the introduced heavy excitation in fig.\ref{fig:Exc} with time.}
\label{fig:fVsEBig} 
\end{figure}

\section{Conclusions}
In our previous works \cite{Michael} we had proposed a solver for the solution of the time dependent Boltzmann scattering integral, with no close-to-equilibrium or fixed population approximations and with realistic band structures and matrix elements. We had further improved our proposed numerical solver to include second order momentum discretisation and adaptive time stepping \cite{1DPaper}. In this work, we showcased the applications of this higher order solver to 2D materials. Using doped graphene as a test  material, we introduced various excitations on the numerical steady state distribution and analysed their thermalisation. Excitations introduced at higher energies thermalised faster than those introduced at lower energies. In case of finite excitations, the population decay towards equilibrium values was observed to be a non-trivial function of time and not a simple exponential decay. We also tested the code for realistic excitations in the same material. Fitting a double exponential function to the decay of the excitation with time, we were able to generate a quantitative insight into the time scales involved in the thermalisation. More importantly, although in the present simulations we used a constant value of scattering amplitude ($w^{e-e}_{0123}=1$), comparing our results to the actual experimental data we can arrive at a relevant value of scattering amplitude for the material thereby fixing the only free parameter in our code and making it truly applicable to a host of novel physical phenomenon concerning 2D materials.

\appendix
\renewcommand\theequation{\Alph{section}.\arabic{equation}}
\renewcommand\thefigure{\Alph{section}.\arabic{figure}}

\section{Basis functions}\label{Appendix:BasisFunctions}
\setcounter{figure}{0}
We need an orthonormal set of polynomials as basis functions. In 1D, we use normalized Lagrange polynomials as the required orthonormal basis set. They have the form,
\begin{equation}
\begin{split}
\Psi_{\substack{B b\\n_0}} (k) &= P_{\substack{B b\\n_0}}(k),  \;\;\; k \in E_{\substack{B \\ n_0}} \\
&=0, \;\;\; otherwise
\end{split}
\end{equation}
where, $E_{\substack{B \\ n_0}}$ denotes the element labelled B in band $n_0$ and $P_{\substack{B b\\n_0}}(k)$ has the form,
\begin{equation}
\begin{split}
&P_{\substack{B 0\\n_0}}= \gamma_{\substack{B 0\\n_0}}\\
& P_{\substack{B 1\\n_0}}=\beta_{\substack{B 1\\n_0}}k+\gamma_{\substack{B 1\\n_0}}\\
& P_{\substack{B 2\\n_0}}=\alpha_{\substack{B 2\\n_0}} k^2+\beta_{\substack{B 2\\n_0}}k+\gamma_{\substack{B 2\\n_0}}
\end{split}
\end{equation}
The six unknown coefficients are determined by imposing the orthonormality condition. 

In 2D, we generate our basis set as products of 1D Lagrange polynomials along the X and Y direction. The 1D Lagrange polynomials along the X and the Y direction are:
\begin{equation*}
    \begin{split}
        \Psi_{0i}=&\frac{1}{\sqrt{\Delta_i}} \\
        \Psi_{1i}=&\frac{2\sqrt{3}}{\sqrt{\Delta_i}} \frac{(k_i-C_i)}{\Delta_i}\\
        \Psi_{2i}=&\frac{\sqrt{5}}{2\sqrt{\Delta_i}} \Bigg(12 \Big(\frac{k_i-C_i}{\Delta_i}\Big)^2-1\Bigg)
    \end{split}
\end{equation*}
Where, i=x or y depending on if the Lagrange polynomials are taken along the X or the Y direction respectively. Accordingly, \{$\Delta_x, \Delta_y$\} are the widths of the mesh elements in the X and Y directions respectively and \{$C_x,C_y$\} are the centers of the mesh elements in the X and the Y directions respectively. Taking the cartesian product of these 1D Lagrange polynomials and limiting the highest order of the resulting polynomial to 2 since we want up to second order basis functions, we get an orthomormal basis set in 2D as,

\begin{equation*}
    \begin{split}
        \Psi_{0}=&\Psi_{0x}*\Psi_{0y}=\frac{1}{\sqrt{\Delta_x \Delta_y}} \\
        \Psi_{1}=&\Psi_{1x}*\Psi_{0y}=\frac{2\sqrt{3}}{\sqrt{\Delta_x\Delta_y}} \frac{(k_x-C_x)}{\Delta_x}\\
        \Psi_{2}=&\Psi_{0x}*\Psi_{1y}=\frac{2\sqrt{3}}{\sqrt{\Delta_x\Delta_y}} \frac{(k_y-C_y)}{\Delta_y}\\
        \Psi_{3}=&\Psi_{1x}*\Psi_{1y}=\frac{12}{\Delta_x^{3/2}\Delta_y^{3/2}} (k_x-C_x)(k_y-C_y)\\
        \Psi_{4}=&\Psi_{2x}*\Psi_{0y}=\frac{\sqrt{5}}{2\sqrt{\Delta_x\Delta_y}} \Bigg(12 \Big(\frac{k_x-C_x}{\Delta_x}\Big)^2-1\Bigg)\\
        \Psi_{5}=&\Psi_{0x}*\Psi_{2y}=\frac{\sqrt{5}}{2\sqrt{\Delta_x\Delta_y}} \Bigg(12 \Big(\frac{k_y-C_y}{\Delta_y}\Big)^2-1\Bigg)
    \end{split}
\end{equation*}

\section{Final form for Integration of the Scattering tensor elements} \label{Appendix:Phi}
The expressions inside the dirac deltas for momentum and energy conservation reduce to simple polynomials, as shown in Eq.~\ref{ScatteringTensor}, following our choice of basis functions. Hence both Dirac deltas can be analytically reduced to obtain a smooth final expression for integration by Monte Carlo method. However, the dissymmetry in the expression of the momentum Dirac delta in Eq.~\ref{ScatteringTensor} would result in different final expressions depending upon the variables on which the deltas are reduced and the order in which the the deltas are inverted. To remove this disparity we effect a mapping of variables as : $k_1 \rightarrow x_A ; k_2 \rightarrow x_B ; k_3 \rightarrow -x_C ; k_4 \rightarrow -x_D +G $. Now Eq.~\ref{ScatteringTensor} takes the form
\begin{widetext}
\begin{equation}\label{PhiSymmetric}
    \begin{split}
\Phi[&F[],l_{Ax},h_{Ax},l_{Ay},h_{Ay},l_{Bx},h_{Bx},l_{By},h_{By},l_{Cx},h_{Cx},l_{Cy},h_{Cy},l_{Dx},h_{Dx},l_{Dy},h_{Dy},\\
&\mu_0,\mu_{A1},\mu_{A2},\mu_{A3},\mu_{A4},\mu_{A5},\mu_{B1},\mu_{B2},\mu_{B3},\mu_{B4},\mu_{B5},\mu_{C1},\mu_{C2},\mu_{C3},\mu_{C4},\mu_{C5},\mu_{D1},\mu_{D2},\mu_{D3},\mu_{D4},\mu_{D5},] \\
=& \int_{\substack{l_{Ax}\\n_0}}^{h_{Ax}}\int_{\substack{l_{Ay}\\n_0}}^{h_{Ay}}dx_Ady_A\int_{\substack{l_{Bx}\\n_1}}^{h_{Bx}}\int_{\substack{l_{By}\\n_1}}^{h_{By}}dx_Bdy_B\int_{\substack{l_{Cx}\\n_2}}^{h_{Cx}}\int_{\substack{l_{Cy}\\n_2}}^{h_{Cy}}dx_Cdy_C\int_{\substack{l_{Dx}\\n_3}}^{h_{Dx}}\int_{\substack{l_{Dy}\\n_3}}^{h_{Dy}}dx_Ddy_D\\
&F[x_A,y_A,x_B,y_B,x_C,y_c,x_D,y_D] \delta(x_A+x_B+x_C+x_D)\;\;\delta(y_A+y_B+y_C+y_D)\\
& \delta[\mu_0+\mu_{A1}x_A+\mu_{A2}y_A+\mu_{A3}x_Ay_A+\mu_{A4}x_A^2+\mu_{A5}y_A^2+\mu_{B1}x_B+\mu_{B2}y_B+\mu_{B3}x_By_B+\mu_{B4}x_B^2+\mu_{B5}y_B^2\\
&+\mu_{C1}x_C+\mu_{C2}y_C+\mu_{C3}x_Cy_C+\mu_{C4}x_C^2+\mu_{C5}y_C^2+\mu_{D1}x_D+\mu_{D2}y_D+\mu_{D3}x_Dy_D+\mu_{D4}x_D^2+\mu_{D5}y_D^2] 
\end{split}
\end{equation}
\end{widetext}
where F[...] incorporates $w^{e-e}_{0123}$ and the basis functions (eg.~$\Psi_{\substack{A a \\n_0}}(k_0)$) in Eq.~\ref{ScatteringTensor}.

Now that we have a symmetric expression in the momentum Dirac delta,we can choose any variables for the reduction of the Dirac deltas. We choose to invert the three Dirac deltas in Eq.~\ref{PhiSymmetric} according to the variables $x_A$, $y_A$ and $x_B$. Reducing $x_A$ and $y_A$ gives,
\begin{align}
	x_A[x_B,x_C,x_D]&=-x_B-x_C-x_D,  \\
	y_A[y_B,y_C,y_D]&=-y_B-y_C-y_D
\end{align}

Substituting for $x_A$ and $y_A$ in the Dirac delta for the energy, we obtain a quadratic equation in $x_B$ as,
\begin{align}\label{Quad}
\delta&(energy)=\\
  &\alpha x_B^2+\beta [y_B,x_c,y_C,x_D,y_D] x_B+\gamma
[y_B,x_c,y_C,x_D,y_D] \nonumber
\end{align}
where,
\begin{widetext}
\begin{align}
\alpha=&\mu_{A4}+\mu_{B4}\\
\beta [y_B,x_c,y_C,x_D,y_D]=&\mu_{B1}-\mu_{A1}+2\mu_{A4}(x_C+x_D)-\mu_{A3}y_A[y_B,y_C,y_D]+\mu_{B3}y_B  \\
\gamma [y_B,x_c,y_C,x_D,y_D]=&\mu_0+\mu_{A2}y_A[y_B,y_C,y_D]+\mu_{A5}(y_A[y_B,y_C,y_D])^2-\mu_{A3}(x_C+x_D)y_A[y_B,y_C,y_D] \\
&+\mu_{B2}y_B +\mu_{B5}y_B^2+2\mu_{A4}x_c x_D+(\mu_{C1}-\mu_{A1})x_C+(\mu_{A4}+\mu_{C4})x_C^2 +\mu_{C2}y_C+\mu_{C5}y_C^2+\mu_{C3}x_Cy_C \nonumber\\
&+(\mu_{D1}-\mu_{A1})x_D+(\mu_{A4}+\mu_{D4})x_D^2 +\mu_{D2}y_D+\mu_{D5}y_D^2+\mu_{D3}x_Dy_D \nonumber
\end{align}
\end{widetext}
Solving Eq.~\ref{Quad} gives two values of $x_B$($x_{B+}$ and $x_{B-}$) which will give two corresponding values of $x_A$ ($x_{A+}$ and $x_{A-}$). Finally we substitute for $x_A$, $y_A$ and $x_B$ in Eq.~\ref{PhiSymmetric} to get
\begin{widetext}
\begin{equation}\label{FinalPhi}
    \begin{split}
\Phi[...]=&\int_{\substack{l_{By}\\n_1}}^{h_{By}}dy_B\int_{\substack{l_{Cx}\\n_2}}^{h_{Cx}}\int_{\substack{l_{Cy}\\n_2}}^{h_{Cy}}dx_Cdy_C\int_{\substack{l_{Dx}\\n_3}}^{h_{Dx}}\int_{\substack{l_{Dy}\\n_3}}^{h_{Dy}}dx_Ddy_D\\
&\Bigg( \frac{F[x_{A+}[y_B,x_C,y_C,x_D,y_D],y_A[y_B,y_C,y_D],x_{B+}[y_B,x_C,y_C,x_D,y_D],y_B,x_C,y_c,x_D,y_D]}{\sqrt{D[y_B,x_C,y_C,x_D,y_D]}}\\
&\Theta_{[l_{Ax},h_{Ax},l_{Ay},h_{Ay}]}[x_{A+}[y_B,x_C,y_C,x_D,y_D],y_A[y_B,y_C,y_D]]\Theta_{[l_{Bx},h_{Bx},l_{By},h_{By}]}[x_{B+}[y_B,x_C,y_C,x_D,y_D],y_B]\\
&\Theta_{[0,\infty]}[D[y_B,x_C,y_C,x_D,y_D]]\\
&+\frac{F[x_{A-}[y_B,x_C,y_C,x_D,y_D],y_A[y_B,y_C,y_D],x_{B-}[y_B,x_C,y_C,x_D,y_D],y_B,x_C,y_c,x_D,y_D]}{\sqrt{D[y_B,x_C,y_C,x_D,y_D]}}\\
&\Theta_{[l_{Ax},h_{Ax},l_{Ay},h_{Ay}]}[x_{A-}[y_B,x_C,y_C,x_D,y_D],y_A[y_B,y_C,y_D]]\Theta_{[l_{Bx},h_{Bx},l_{By},h_{By}]}[x_{B-}[y_B,x_C,y_C,x_D,y_D],y_B]\\
&\Theta_{[0,\infty]}[D[y_B,x_C,y_C,x_D,y_D]]\Bigg)
\end{split}
\end{equation}
\end{widetext}

Here, $\Theta[...]$ is the heaviside function between the limits of the reduced variables (In this case,$x_A$, $y_A$ and $x_B$) and D is the determinant of the quadratic equation in Eq.~\ref{Quad} 
\begin{equation}
    D=(\beta [y_B,x_c,y_C,x_D,y_D])^2-4\alpha\gamma [y_B,x_c,y_C,x_D,y_D]
\end{equation}
The expression in Eq.~\ref{FinalPhi} does not contain any Dirac deltas and has been reduced from a 8 dimensional integral to a 5 dimensional integral. We use standard Monte Carlo integration method to obtain the integration value, yet other quadrature methods can be used as well.

\section{Structure of the chosen functional form of the Scattering Amplitude}\label{Appendix:W}

In this section we detail the theoretical basis for the restrictions on the functional form of the scattering amplitude used in this study. We used the simplest approximation of the scattering amplitude, i.~e.~a constant. This choice is obviously an oversimplification of reality, and an appropriate dependence needs to be obtained and used as input for our numerical method. In spite of that, the choice of a constant scattering amplitude, (with the constant fitted to the experiments) allows for the semiquantitative understanding of a number of thermalisation timescales, as often a very strong momentum or energy dependence of the dynamics is due to the phase space (which is treated without approximations) rather than the matrix elements. 

There is only one case where using a constant approximation leads to problems that must be addressed: in the limit of momentum transferred during the scattering going to zero for scatterings within the same band. In this limit, the actual change of the quasiparticle populations tend to vanish, as initial and final states tend to become the same. Notice that this is true regardless of the scattering amplitude. Nonetheless there is a numerical problem arising: since the joint density of states of the transition diverges in this limit, numerical errors tend to grow, and become bigger than the actual population change.

We rectify this by introducing the following functional form of the scattering amplitude,

\begin{equation}
w^{e-e}_{0123}=
\begin{cases}
\sqrt{\frac{\Delta_k}{\Delta_{mesh}}}, & \Delta_k \leq \Delta_{mesh}\\
1,              & \text{otherwise}
\end{cases}
\end{equation}

where $\Delta_k$ is the net transferred momentum, i.e. $|\vec{\bold{k}}_1+\vec{\bold{k}}_2-\vec{\bold{k}}_3-\vec{\bold{k}}_4|$ and $\Delta_{mesh}$ is the diagonal width of the rectangular mesh element i.e. $\Delta_{mesh}=\sqrt{\Delta_x^2+\Delta_y^2}$, where $\Delta_x$,$
\Delta_y$ are the widths of the mesh element in the x and the y direction respectively as before. Since we use a uniform mesh the value of $\Delta_{mesh}$ remains fixed once the mesh is defined. Such choice of the functional form of the scattering amplitude ensures that even if the joint density of states diverge the population change tends to zero as expected. 

Below, we show in a systematic way why the scattering matrix element tends to 0 for electron-electron scatterings where the transferred momentum tends to vanish.

Consider a four leg electron-electron scattering process. We assume here that all the involved electronic states in the scattering belong to the same band and so they are labelled by only their respective momenta, $\{k_1,k_2,k_3,k_4\}$. 

Let $\hat N_{k}$ denote the number operator for the instantaneous occupation number of a state labelled by k. We remind that the expectation value of this number operator, $<\hat N_k>$, becomes the population distribution function, $f(t,k)$, in the context of the Boltzmann Equation.

Using the Quantum Fokker-Planck equation it can be shown that the change of $<\hat N_k>$ with time goes as:

\begin{equation} \label{Nk}
    \frac{d<\hat N_k>}{dt} \sim [\hat N_k, V_{int}]
\end{equation}

where $V_{int}$ is the number conserving interaction term for the scattering and it is given as,
  \begin{equation*}
      V_{int}=\frac{1}{2} \sum_{k_1,k_2,k_3} U_{k_1,k_2,k_3,k_4} a^{\dagger}_{k_4} a^{\dagger}_{k_3} a_{k_2} a_{k_1}
  \end{equation*}
$[\hat N_k, V_{int}]$ is the commutator of the number operator with the interaction term. Eq.\ref{Nk} implies that when the number operator corresponding to a particular state commutes with the interaction term there is no change in the occupation of that state.

For simplicity of calculation, let us consider the change in the occupation number of the state labelled by $k_4$. The corresponding commutator $[\hat N_{k_4}, V_{int}]$, can be derived as follows:

\begin{widetext}
\begin{equation*}
     \begin{split}
         \hat N_{k4} a_{k4}^\dagger a_{k3}^\dagger a_{k2} a_{k1} 
         =& (a_{k4}^\dagger+a_{k4}^\dagger \hat N_{k4}) a_{k3}^\dagger a_{k2} a_{k1} \\
         =&a_{k4}^\dagger a_{k3}^\dagger a_{k2} a_{k1} +a_{k4}^\dagger (a_{k4}^\dagger \delta_{k3k_4}+a_{k3}^\dagger \hat N_{k4}) a_{k2} a_{k1}\\
         =&a_{k4}^\dagger a_{k3}^\dagger a_{k2} a_{k1} + a_{k4}^\dagger a_{k4}^\dagger \delta_{k3k_4} a_{k2} a_{k1}+a_{k4}^\dagger a_{k3}^\dagger(-a_{k4} \delta_{k2k_4}+a_{k2}\hat N_{k4}) a_{k1}\\
         =&a_{k4}^\dagger a_{k3}^\dagger a_{k2} a_{k1} + a_{k4}^\dagger a_{k4}^\dagger \delta_{k3k_4} a_{k2} a_{k1}-a_{k4}^\dagger a_{k3}^\dagger a_{k4} \delta_{k2k_4}a_{k1}+a_{k4}^\dagger a_{k3}^\dagger a_{k2}(-a_{k4} \delta_{k1k_4}+a_{k1}\hat N_{k4})\\
         =&a_{k4}^\dagger a_{k3}^\dagger a_{k2} a_{k1} + a_{k4}^\dagger a_{k4}^\dagger \delta_{k3k_4} a_{k2} a_{k1}-a_{k4}^\dagger a_{k3}^\dagger a_{k4} \delta_{k2k_4}a_{k1}-a_{k4}^\dagger a_{k3}^\dagger a_{k2}a_{k4} \delta_{k1k_4}\\
         +&a_{k4}^\dagger a_{k3}^\dagger a_{k2} a_{k1}\hat N_{k4}
     \end{split}
 \end{equation*}
 So,
 \begin{equation}\label{Comm}
     [\hat N_{k4},\hat V_{int}]=a_{k4}^\dagger a_{k3}^\dagger a_{k2} a_{k1} + a_{k4}^\dagger a_{k4}^\dagger \delta_{k3k_4} a_{k2} a_{k1}-a_{k4}^\dagger a_{k3}^\dagger a_{k4} \delta_{k2k_4}a_{k1}-a_{k4}^\dagger a_{k3}^\dagger a_{k2}a_{k4} \delta_{k1k_4}
 \end{equation}
\end{widetext}

Eq.\ref{Comm} presents two scenarios when the commutator is zero or in other words $\frac{d<\hat N_k>}{dt}=0$.

i) $k_4=k_1$ Or $k_4=k_2$

ii) $k_4=k_1$ and $k_4=k_2$ and $k_4=k_3$ i.e. all the involved states are the same. This is a trivial case and we do not pursue it further.

In case of scenario (i), the momentum Dirac delta dictates that the remaining two 'k's must also be equal to each other and hence the net transferred momentum is zero. Or equivalently, when the net transferred momentum is zero the number operator commutes with the interaction term and so the change in the population distribution function must be zero.

 \section{Scattering Rates}\label{Appendix:ScatRates}
 
Eq.~\eqref{ScatteringIntegral} gives the change of the particular population distribution function \( f_{n_0} (t,\vec{\bold{k}}_0) \) in a scattering process. 
It can be proved \cite{Michael} that for a small and localized excitation in the state \(|n_0\vec{\bold{k}}_0>\) over the equilibrium distribution (appropriate Fermi-Dirac distribution in our case), the population decays back exponentially to the equilibrium distribution. The inverse of the time constant of this exponential decay is the k-resolved scattering rate, $\lambda_{n_0}(\vec{\bold{k}}_0)$ which is given as (see \cite{Michael} for details),

\begin{widetext}
\begin{equation}\label{ScatteringRates}
\begin{split}
\lambda_{n_0}(\vec{\bold{k}}_0)= \sum_\bold{G} \int \int \int_{V^3_{BZ}} & d^2\vec{\bold{k}}_1 \; d^2\vec{\bold{k}}_2 \; d^2\vec{\bold{k}}_3 \; \; w^{e-e}_{0123} \;\; \;\delta(\epsilon_{n_0}(\vec{\bold{k}}_0)+\epsilon_{n_1}(\vec{\bold{k}}_1)-\epsilon_{n_2}(\vec{\bold{k}}_2)-\epsilon_{n_3}(\vec{\bold{k}}_3)) \\ & \delta(\vec{\bold{k}}_0+\vec{\bold{k}}_1-\vec{\bold{k}}_2-\vec{\bold{k}}_3+\bold{G})[(1-f_{1}^{eq})f_{2}^{eq} f_{3}^{eq}-f_{1}^{eq}(1-f_{2}^{eq})(1-f_{3}^{eq})]
\end{split}
\end{equation}
\end{widetext}

where $f_{i}^{eq}$ ( shorthand for $f_{n_i}^{eq}(\vec{\bold{k}}_i)$) is the relevant equilibrium distribution (Fermi-Dirac distribution in this case). Eq.~\ref{ScatteringRates} appears very similar in structure to Eq.~\ref{ScatteringIntegral}. But there is a very crucial difference. Notice that the phase factor in Eq.~\ref{ScatteringRates} is composed of equilibrium population distributions instead of the time dependent population distributions. Therefore, the integral in eq.\ref{ScatteringRates} is not a quartic operator and, as we will show in the next section, it can be estimated at a fraction of the computational cost of the complete time propagation in Eq.~\ref{ScatteringIntegral}. 

\section{Discretized form of Scattering rates}

To obtain the discretised form of the scattering rates,$\lambda_{\substack{ A a' \\n_0}}$, we project the equation for the scattering rates, Eq.~\ref{ScatteringRates}, on the chosen basis functions and follow a similar procedure as above (see ref.\cite{Michael} for details) to obtain:

\begin{widetext}
\begin{equation}\label{DescritizedScatRates}
\begin{split}
\lambda_{\substack{A a' \\n_0}}=\sum_{A a} \sum_{B b} \sum_{C c} \sum_{D d} S^{a' a b c d}_{\substack{A A B C D \\ n_0 n_1 n_2 n_3}}  \Bigg( & 1_{\substack{A a \\ n_0}} \Big[1_{\substack{B b \\ n_1}}- f_{\substack{B b \\n_1}}^{eq}\Big]  f_{\substack{C c \\n_2}}^{eq} f_{\substack{D d \\n_3}}^{eq} -
 1_{\substack{A a \\ n_0}} f_{\substack{B b \\n_1}}^{eq} \Big[1_{\substack{C c \\ n_2}}-f_{\substack{C c \\n_2}}^{eq}\Big] \Big[1_{\substack{D d \\ n_3}}-f_{\substack{D d \\n_3}}^{eq}\Big] \Bigg) \\
\end{split}
\end{equation}
\end{widetext}

\bibliographystyle{unsrt}
\bibliography{References}

\end{document}